\documentclass[12pt,amsmath,amssymb,showpacs,aps,floats,nofootinbib,preprint, superscriptaddress]{revtex4}
\usepackage{graphicx}
\usepackage{verbatim}
\usepackage{ulem}
\usepackage{color}
\usepackage{epsfig}
\usepackage{epstopdf}
\usepackage{dcolumn}
\usepackage{bm}
\usepackage{amsmath}
\usepackage{amsfonts}
\usepackage{amssymb}
\usepackage{graphicx}
\usepackage{latexsym}
\usepackage{rotating}
\usepackage{multirow}
\usepackage{slashed}
\usepackage{xcolor}
\usepackage{hyperref}
\usepackage{natbib}
\def\be{\begin{equation}}
\def\ee{\end{equation}}

\newcommand{\bea}{\begin{eqnarray}}
\newcommand{\eea}{\end{eqnarray}}
\newcommand{\bwt}{\begin{widetext}}
\newcommand{\ewt}{\end{widetext}}

\def\u
\def\hZ{\widehat Z}

\def\eed{\end{document}}

\def\m_z{m_{\textrm {Z}}}

\renewcommand{\u}{\rm{u}}

\def\be{\beta}

\newcommand{\wt}{\widetilde}

\def\rm#1{\textrm{#1}}

\def\ra{\rightarrow}

\def\f{\frac}

\def\bwt{\begin{widetext}}
\def\ewt{\end{widetext}}
\def\be{\begin{equation}}
\def\ee{\end{equation}}
\def\bea{\begin{eqnarray}}
\def\eea{\end{eqnarray}}
\def\bean{\begin{eqnarray*}}
\def\eean{\end{eqnarray*}}
\def\bary{\begin{array}}
\def\eary{\end{array}}
\def\bit{\begin{itemize}}
\def\eit{\end{itemize}}

\def\ra{\rightarrow}

\def\su5u1{SU(5) \times U(1)}
\def\fsu5u1{SU(5) \times U(1)'}
\def\so10{SO(10)}
\def\sq20{SO(10) \times SO(10)}

\DeclareMathOperator*{\minmt}{min}

\def\ra{\rightarrow}

\def\f{\frac}

\def\L{\left(}
\def\R{\right)}

\def\ra{\rightarrow}

\def\su5u1{SU(5) \times U(1)}
\def\fsu5u1{SU(5) \times U(1)'}
\def\so10{SO(10)}
\def\sq20{SO(10) \times SO(10)}

\begin{document}
\title{Simplified Supersymmetry with Sneutrino LSP at 8 TeV LHC}
\author{Jun Guo}
\email{hustgj@itp.ac.cn}
\affiliation{State Key Laboratory of Theoretical Physics, Institute
of Theoretical Physics, Chinese Academy of Sciences, Beijing 100190,
P. R. China}
\author{Zhaofeng Kang}
\email{zhaofengkang@gmail.com}
\affiliation{Center for High-Energy Physics, Peking University,
Beijing, 100871, P. R. China}
\author{Jinmian Li}
\email{jmli@itp.ac.cn}
\affiliation{State Key Laboratory of Theoretical Physics, Institute
of Theoretical Physics, Chinese Academy of Sciences, Beijing 100190,
P. R. China}
\author{Tianjun Li}
\email{tli@itp.ac.cn}
\affiliation{State Key Laboratory of Theoretical Physics, Institute
of Theoretical Physics, Chinese Academy of Sciences, Beijing 100190,
P. R. China}
\affiliation{School of Physical Electronics, University of
Electronic Science and Technology of China, Chengdu 610054, P. R.
China}
\author{Yandong Liu}
\email{ydliu@itp.ac.cn} \affiliation{State Key Laboratory of
Theoretical Physics, Institute of Theoretical Physics, Chinese
Academy of Sciences, Beijing 100190, P. R. China}

\begin{abstract}

The current searches of supersymmetry (SUSY) are based on the
neutralino lightest sparticle (LSP). In this article we instead
focus on SUSY with sneutrino LSP.
It is well motivated in many contexts, especially in which sneutrino services as a dark matter candidate. 
We first develop a
simplified model, which contains the stop, chagino/neutralino and
sneutrino, to describe the LHC phenomenologies of a large
class of models with sneutrino LSP. Then we investigate bounds on
the model using the SUSY searches at the 8 TeV LHC. Strong exclusion
limits are derived, e.g., masses of stop and chargino can be
excluded up to about 900 GeV and 550 GeV, respectively. We also
propose optimizations for some searches without turning to higher
energy and luminosity.


\end{abstract}

\maketitle

\newpage

\section{Introduction and motivation}

The existence of dark matter (DM) is commonly accepted by virtue of many
convincing gravitational evidences~\cite{Bertone:2004pz}. It is then
regarded as a strong hint for new physics beyond the standard model
(SM). Interestingly, once the R-parity conservation is imposed, the
supersymmetric standard models (SSMs), originally proposed to solve
the notorious gauge hierarchy problem, will naturally provide DM
candidates, i.e., the lightest supersymmetric particle (LSP). Hence, the LSP dark matter candidates receive intensive attention.

In the minimal SSM (MSSM), there are three LSP DM candidates, the
lightest neutrilino $\chi_1$, gravitino and the lightest
(must be left-handed) sneutrino $\tilde{\nu}_1$. Among them,
$\chi_1$, a typical weakly interacting massive particle
(WIMP), is a good candidate~\cite{Jungman:1995df}. While
$\tilde{\nu}_1$, despite also being a WIMP, is not a that good DM. The
reasons are two-folded. First, it possesses full $SU(2)_L$ gauge
interaction, so it annihilates fast, rendering its correct relic
density achieved only for quite heavy $\tilde{\nu}_1$ (a few TeVs).
Second, the $Z-$boson mediated DM-nucleon spin-independent (SI)
scattering has a too large cross section
$\sigma_{\rm{SI}}$~\cite{Falk:1994es} such that it has been
definitely excluded by the present DM direct detection experiments
like LUX~\cite{Akerib:2013tjd}. However, the situation may change
drastically when we go beyond the MSSM following another guideline
of new physics, the tiny but non-vanishing neutrino masses.

Seesaw mechanisms provide good avenues to understand the origin of neutrino mass~\cite{Yanagida:1979as,GellMann:1980vs,PhysRevD.34.1642}. In them, we may introduce extra SM singlets, collectively denoted as
$N$. They carry lepton number and their superpartners $\wt N$, along with $\wt \nu_L$, constitute the extended sneutrino system. Now, given a low seesaw scale, the lightest sneutrino can be a viable WIMP DM candidate. Typically, it acquires correct thermal relic density $\Omega_{\rm DM}h^2\simeq0.1$ in two cases. One is that the sneutrino LSP is a well mixture of left-handed sneutrino $\wt \nu_L$ and $\wt N$, and then the proper $\wt \nu_L$ component helps to reduce the number density of sneutrino~\cite{ArkaniHamed:2000bq,Arina:2008bb,MarchRussell:2009aq,Belanger:2010cd}. A good case in point is in the supersymmetric low-scale type-I seesaw models which includes a large soft trilinear term $A_\nu \wt L H_u\wt N$ ($A_\nu$ is not suppressed by the small Yukawa coupling). The other one is that the $\wt N-$like LSP has significant couplings to the Higgs superfields~\cite{Arina:2008bb,Kang:2011wb,An:2011uq,
DeRomeri:2012qd,BhupalDev:2012ru,Choi:2013fva,Guo:2013sna,Banerjee:2013fga} such as
in the supersymmetric inverse seesaw models, in which a supersymmetric term $y_\nu L H_u N$ with $y_\nu\sim {\cal O}(0.1)$ is allowed. In summary, a thermal sneutrino LSP means that $\wt N$ does not decouple. In particular, the LSP is expected to have a sizable coupling either to wino (via its $\wt \nu_L$ component) or Higgsino (via a large Yukawa coupling)~\footnote{Of course, there are exceptions, say $\wt N$ has  new gauge/Yukawa interactions~\cite{Cerdeno:2009dv,Lee:2011ti,Bandyopadhyay:2011qm,Hirsch:2012kv,DeRomeri:2012qd,Deppisch:2008bp}.}.

This article is devoted to investigate the status of SUSY with a sneutrino LSP at the 8 TeV LHC. From the above arguments, the ordinary sparticles cascade decay into the LSP are prompt~\footnote{ The resulted signatures are contrast to the signatures of long-lived charged sparticle or displaced vertex, which usually appear in the case of quite weak couplings between the sneutrino LSP and sparticles in the visible
sector~\cite{Batell:2013bka,Cerdeno:2013oya,Arina:2013zca}.}. Moreover, in general the sneutrino LSP leads to multi-leptons in the final states of these decays. Compared to the conventional neutralino LSP, it is thus more likely to expose SUSY at a hadronic collider. For example, based on the supersymmetric inverse seesaw, Ref.~\cite{Mondal:2012jv} studied the
tri-lepton plus missing $E_T$ signature from $C^{\pm}_1
\chi_2$ production. Again in this model,
Ref.~\cite{BhupalDev:2012ru} studied the signature of same sign
dilepton plus jets and missing energy from gluino- and squark- pair
as well as the squark-gluino associated productions.  Other relevant studies, some of which studied the previous signatures earlier, can be found in~\cite{Thomas:2007bu,Katz:2009qx,Lee:2011ti,Belanger:2011ny,deGouvea:2006wd,Cheung:2011ph,Das:2012ze}. But
these remarkable signatures are fairly model dependent thus
not representative signatures of SUSY with a sneutrino LSP. Based on the simplified model for this SUSY scenario, we find that the opposite-sign dilepton plus missing energy with or without $b$-jets do capture its most common collider feature. And the current SUSY searches have imposed stringent constraints on it.

The paper is organised as following. In Section II, we will develop the simplified model to describe a large class of supersymmetric models with a sneutrino LSP. We also discuss the decay topologies in the model. In Section III we make the collider setup and investigate the bounds on the model in light of the current LHC searches. Possible optimizations are also briefly discussed. Section IV is the discussion and conclusion. Some necessary and complementary details are given in the Appendices.



\section{Simplified model for SUSY with sneutrino LSP}

In this section we will first develop the simplified model for SUSY with sneutrino LSP and introduce the conventions. Then, we analyze the basic collider features of the model, including decay lifetime and topologies of the sparticles within the model.

\subsection{Simplified model}

We are at the position to make a simplified model for SUSY with sneutrino LSP, $\wt \nu$. On top of $\wt \nu$, the model should contain three other superpartners, the stop $\wt t$,  chargino $C^\pm$ and  neutralino $\chi$. They represent the colored sparticles and electroweak sparticles, respectively. In particular, $C/\chi$ are always relevant when we are studying the colored sparticle (like stop) decaying into the sneutrino LSP. In addition, they have been extensively searched at the present LHC (based on the ordinary SUSY with neutralino LSP) and thus we can avail ourself of the public data to constrain them here. In a word, they are the proper minimal field content for the simplified model. The relevant interactions are casted in the following effective Lagrangian
\begin{align}\label{simp}
-{\cal L}_{eff}=&m_{\wt t}^2|\wt t|^2+m_{\wt\nu}^2|\wt\nu|^2+m_C\bar
C C+\f{1}{2}m_\chi\chi\chi\cr &+ \left[\bar C^-
 \L g_L^b P_L +g_R^b P_R
\R b^- \wt t^- +\bar C^- \L g_L^e P_L +g_R^e P_R \R e^- \,\wt \nu
+h.c.\right]\cr & +\left[\bar \chi
 \L g_L^t P_L +g_R^t P_R
\R t \wt t^* +\bar \chi
 \L g_L^\nu P_L +g_R^\nu P_R
\R \nu \wt \nu +h.c.\right],
\end{align}
with flavor index of leptons implied. Couplings involving the gauge
bosons are not written out explicitly, and they are relevant only
when we are discussing the chargino/neutralino productions. The sneutrino is assumed to be complex, but sneutrino being real scalar will not affect our discussions much.

The lagrangian of the simplified model is not simple. It contains quite a few free parameters. The effective coupling constants $g_{L/R}$ can be expressed in terms of the gauge/Yukawa coupling constants and the mixing angles in the stop and chargino sectors, etc. For instance,
in the supersymmetric inverse seesaw models the sneutrino LSP DM may be a highly complex scalar, thus dominated by singlets~\cite{Guo:2013sna}. Then we can further simplify the
model by working in the Higgsino-limit with $C^-=(\wt H_d^-,(\wt
H_u^+)^\dagger)^T$, where the couplings can be derived from the term in the superpotential, $y_\nu L H_u N$:
\begin{align}
\quad g_L^e=1,\quad g_R^e=0; \quad g_L^\nu=1,\quad g_R^\nu=0.
\end{align}
Similar chiral structure appears in the wino-limit. In the collider
search the concrete values of coupling
constants are not important, except that branching ratios are of concern. Thus in
various chiral limits we simply set the corresponding $g_{L/R}$ to
be 1 or 0. Different chiral structures will lead to similar results, if
we do not rely on the angular distributions of final states. We have more comments on this point in Appendix~\ref{vali}.

To end up this section, we would like to add some comments on the
applicability of the simplified model in Eq.~(\ref{simp}). Firstly, in
the non-thermal gravitino dark matter scenario, where gravitino
gains correct relic density via the left-handed sneutrino NLSP later
decay~\cite{Feng:2003xh,Feng:2004mt,Kang:2011ny}, sneutrino actually behaves as the LSP at colliders.
Thus, such a SUSY scenario can be described by the simplified model. Next, in fact we do not need so large $g^{e,\nu}$ to make the sneutrino DM thermal. We only require it to ensure chargino dominantly decays into
$\wt \nu+e$ rather than into $\chi+W^*$. In this sense non-thermal sneutrino DM~\cite{Asaka:2005cn,Gopalakrishna:2006kr} also may be described. Finally, $g^t$, etc., should be sufficiently large so that $\wt t/C/\chi$ decay promptly  at the LHC. 

\begin{figure}[htb]
\begin{center}
\includegraphics[width=0.48\columnwidth]{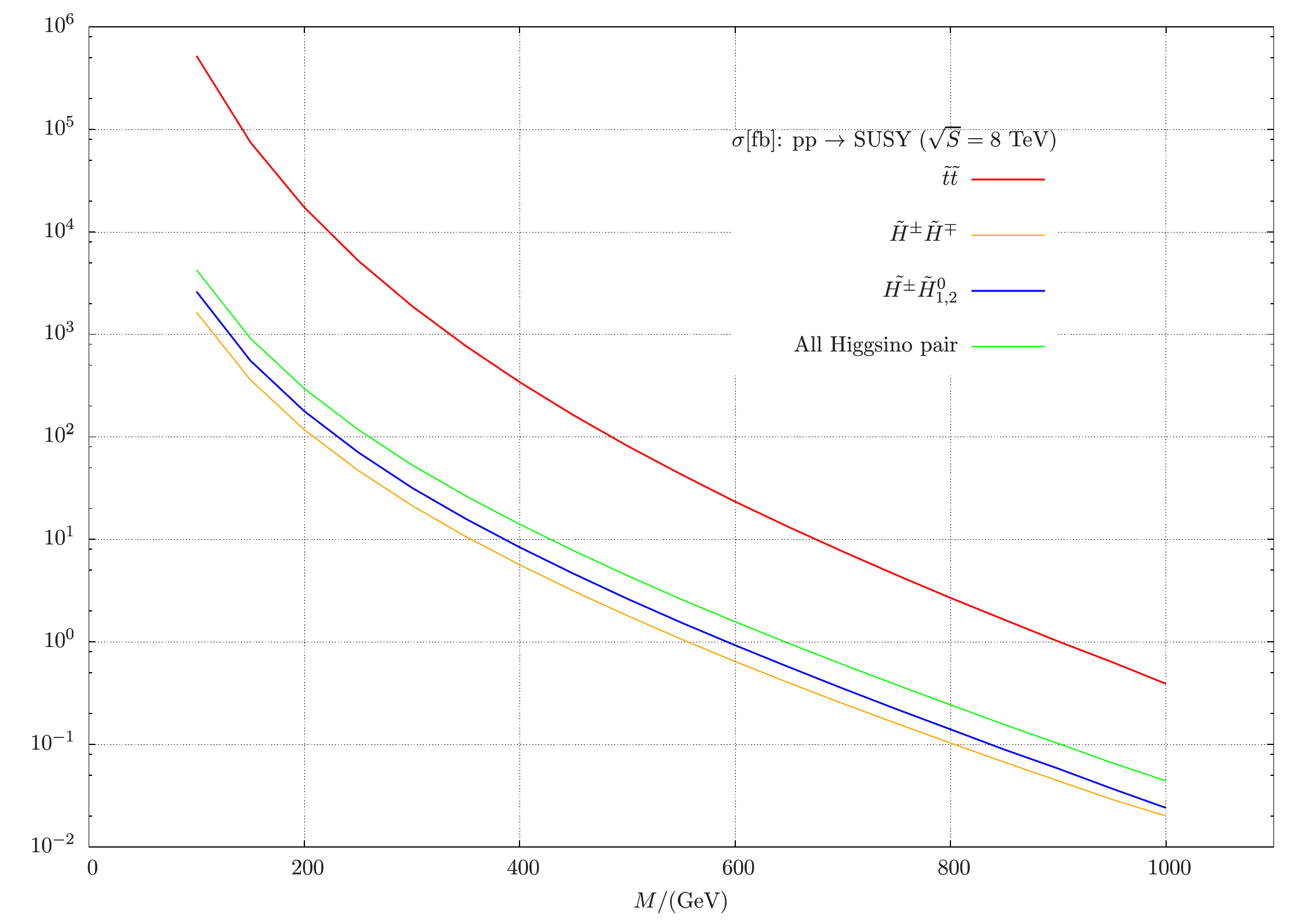}
\includegraphics[width=0.48\columnwidth]{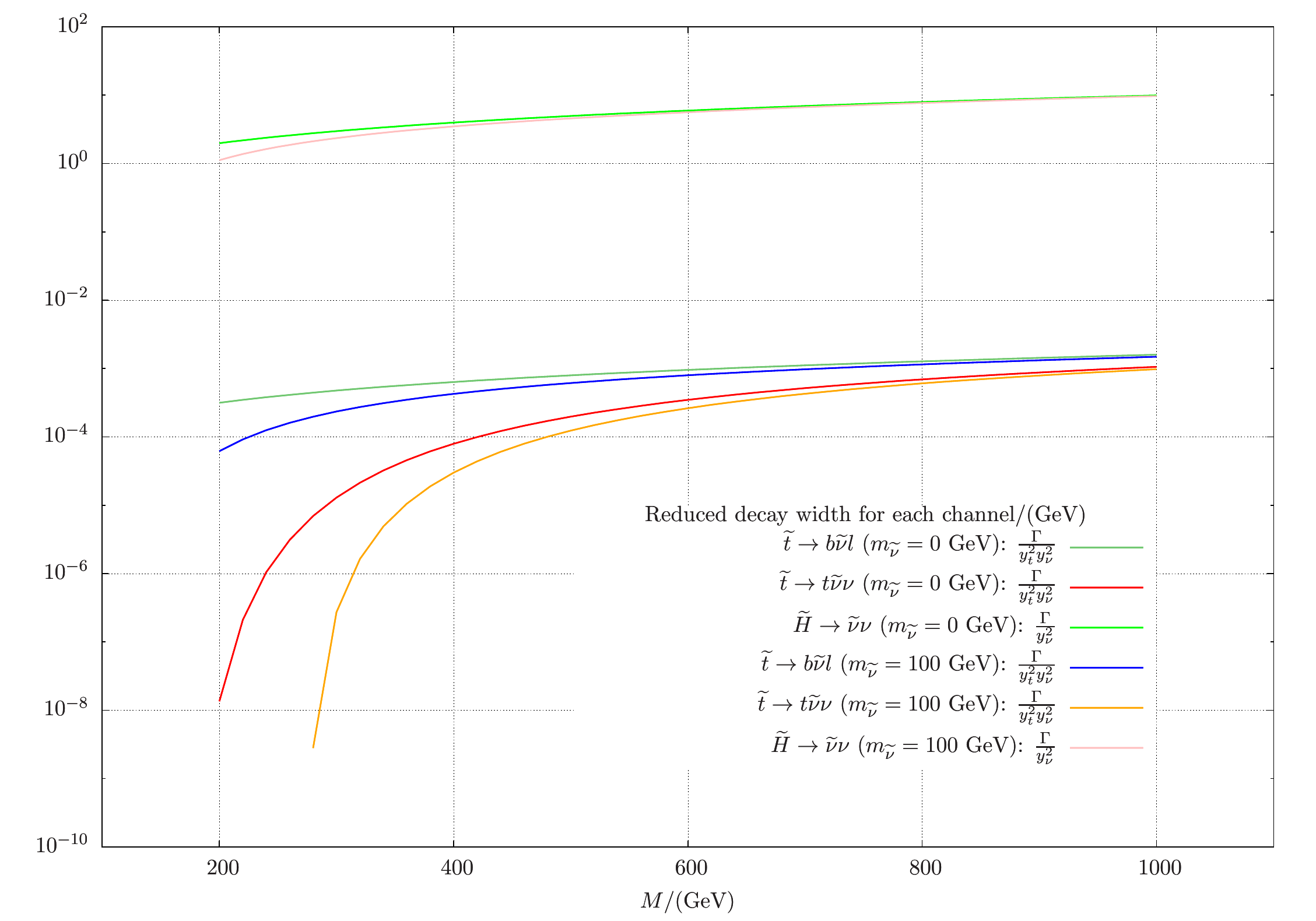}
\caption{Left: Production cross sections of stop pair,
$\tilde{H}^\pm\tilde{H}^0$ and $\tilde{H}^\pm\tilde{H}^\mp$. Right:
Decays of stop/Higgsino. In top3 and top4, we set $m_\chi=1.1m_{\wt t}$ and
$m_C=1.5m_{\wt t}$, respectively. \label{xsec}}
\end{center}
\end{figure}

To examine the second and third points aforementioned, we perform analytical calculations of the relevant decay rates and cast the complicated analytical expressions in Appendix~\ref{adecay}. For illustration, we consider the right-handed stop/sneutrino and Higgsino-limit. First, to make chargino dominantly decays into sneutrino, one may need
\begin{align}
y_{\nu}\gtrsim \sqrt{ \frac{32\pi\Gamma_{C\rightarrow \chi W^*}}{
(m_{\widetilde{\chi}}^2-m_{\widetilde{\nu}}^2)^2/{ m_{\widetilde{\chi}}^3}}}.
\end{align}
For example, for $m_{\widetilde{\nu}}=100$ GeV and $m_{\chi}=0.99m_{C}$, we need $y_{\nu}\gtrsim 10^{-5}$;
While as the degeneracy decrease slightly, says $m_{\chi}=0.95m_{C}$, we need a much larger $y_{\nu}\gtrsim 10^{-3}$. Second, the traveling distances of stop, etc., are estimated as:
\begin{equation}
c\tau = \left\{
\begin{array}{rl}
\frac{1.975\times 10^{-16}}{\Gamma_R} y_t^2 y_\nu^2, & \tilde{t}\to t \nu \tilde{\nu},\,\,b l \tilde{\nu},\\
\frac{1.975\times 10^{-16}}{\Gamma_R} y_\nu^2, & \tilde{H}\to \nu
\tilde{\nu}.
\end{array} \right.
\end{equation}
From Fig.~\ref{xsec} we see that as long as $y_\nu \gtrsim
10^{-5}$ ($y_\nu\gtrsim 10^{-7}$), stop (chargino) will three-body
(two-body) decay promptly at the LHC.
\label{proddecay}

\subsection{Decay topologies in the simplified model}

In the simplified model, the decay chains of $\wt t$, etc., terminate at the sneutrino LSP. Along these chains with chargino propagator, charged leptons are produced and potential to present characteristic signatures at the LHC. While decays with neutralino propagator bring no substantial difference to the conventional SUSY.

Before heading towards the main intent of this subsection, analyzing
the decay topologies of $\wt t$ and $C,\,\chi$, we first show their
production cross sections at the LHC. Stop, a colored sparticle, has
a large cross section of pair production. While productions of the
electroweak sparticles $C/\chi$ depend on their ingredients. If
$\chi$ is a $SU(2)_L$ singlet, such as bino or singlino in the
NMSSM, its direct production will be suppressed. However, when $C/\chi$
are in the Higgsino-limit or wino-limit, the $C$ and $\chi$
associated production will dominate over others. We show the
numerical results, calculated by
Prospino2~\cite{Beenakker:1996ed,Beenakker:1997ut,Beenakker:1999xh},
in Fig~\ref{xsec}.

\begin{figure}[htb]
\begin{center}
\includegraphics[width=0.3\columnwidth]{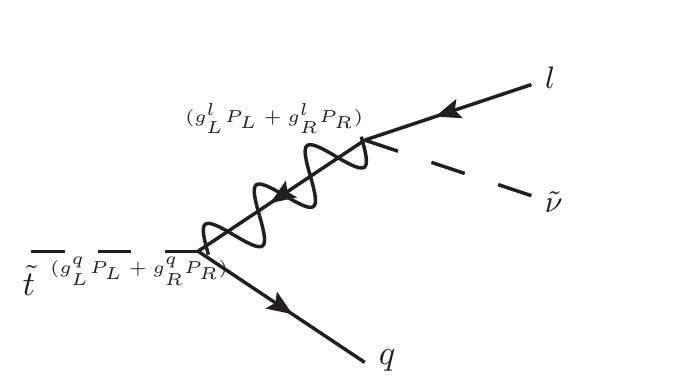}
\includegraphics[width=0.3\columnwidth]{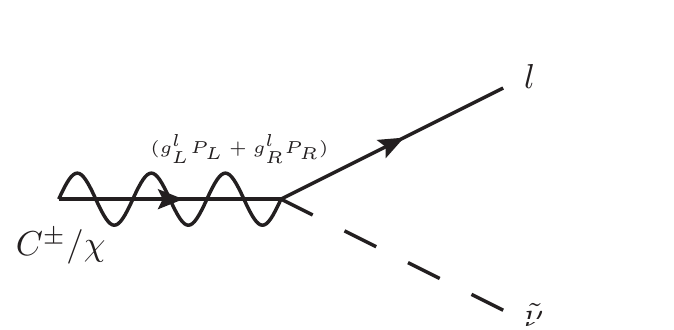}
\includegraphics[width=0.3\columnwidth]{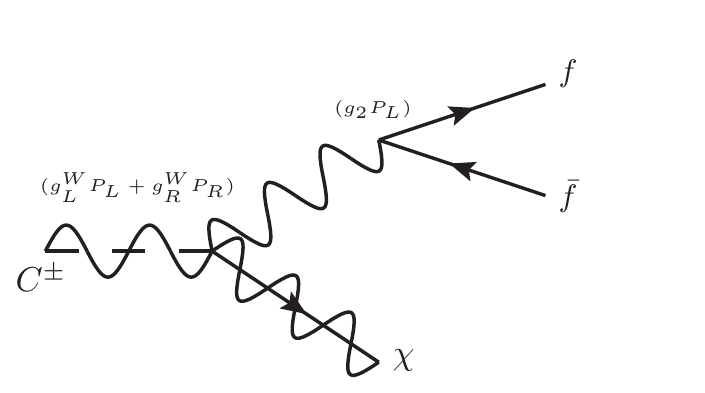}
\caption{Decay Processes.
\label{procs}}
\end{center}
\end{figure}

We now proceed to discuss the decay topologies. Evidently, the
complete list is hard to be exhausted, because the decays heavily depend on the mass hierarchies. We thus only concentrate on the typical
scenarios, which can be seen in the Feymann diagrams
Fig.~\ref{procs}. If stop is so heavy, says owing to the
relatively heavy SM-like Higgs boson, that it decouples from the
simplified model, then we will have to probe SUSY in this scenario via
the $C/\chi$ productions and decays (into sneutrino). Three decay
topologies are generated
\begin{align}\label{}
pp&\ra C\bar C\ra l^\pm l^\mp+E^{\text{miss}}_T,\\
&\ra C^\pm \chi\ra l^\pm +E^{\text{miss}}_T,\\
&\ra \chi\chi\ra E^{\text{miss}}_T.
\end{align}
The present LHC searches are only sensitive to the first topology as will be shown later.
If stop is relatively light, it will have a large production rate and
give additional visible particles (such as the top or bottom quark)
along the decay chains mediated by $C/\chi$. Basically, the stop
decay topologies are classified into four categories:
\begin{description}
  \item[Top1] Neutralino is the next-to LSP (NLSP), with chargino decoupled. Moreover, stop is heavy enough so that $\chi$ is on-shell. Evidently, we can not distinguish this case to the ordinary case, i.e., the stop pair production with $\wt t\ra t+ E^{\text{miss}}_T$.
  \item[Top2] Stop instead is the NLSP, and then $\chi$ is off-shell, i.e., the decay is three-body which means that the top quark of top2 is softened relative to that of top1. As a consequence, it is more difficult to hunt the stop than in the ordinary scenario.
       \item[Top3$\&$4] Corresponding to top1$\&$2, these two  are got by replacing $\chi$ with $C$. By virtue of the hard leptons in the final states, top3$\&$4 have better discovery prospects.
\end{description}
Based on these decay topologies, in the following section we will
derive the current bounds on the simplified model by explaining the
relevant SUSY searches at the 8 TeV LHC.

\section{Constraining the simplified model at 8 TeV LHC}

In this section we derive the bounds on chargino/stop in terms of
the decay topologies given previously. We also make comments on the possible optimizations of some searches for the corresponding signatures.

\subsection{Event Generation and Testing Procedures}

In this paper events are generated by
MadGraph5~\cite{Alwall:2011uj}, where
Pythia6~\cite{Sjostrand:2006za} and PGS~\cite{pgs4} have been packed
to implement parton shower, hadronization and detector simulation.
For detector simulation with PGS, we take the $b$-tagging efficiency
70\%, with $c$-mistag and light-jet mis-tag rates 20\% and 10\%,
respectively. Efficiency of tau-tagging is involved~\cite{ATLAS-CONF-2011-152}, and we simply take the ``medium" jet discrimination quality for tau-tagging, whose efficiency is 40$\%$ (We neglect the mis-tag rate for non-$\tau$ jets, which is quite low, $\sim1\%$). For later convenience, we list the tau decay branching ratios:
\begin{align}
&{\rm {Br}}(\tau \to e \nu_e \nu_\tau)=17.9\%,\quad {\rm {Br}}(\tau \to \mu \nu_\mu \nu_\tau)=17.4\%,\cr
&{\rm {Br}}(\rm {hadronic 1-prong decay})=49.5\%,\quad {\rm {Br}}(\rm {hadronic 3-prong decay})=15.2\%.
\end{align}

The current LHC searches that have sensitivity to the decay topologies in the simplified model are shown in Table~\ref{sLHC}. In the simulation we only consider one specific chirality structure, e.g., $\tilde{t}_R \to b_L \bar e_L \tilde{\nu}_R$ (All decays are prompt and have unit branching ratio). We have explicitly checked that changing helicities can affect the distribution of events merely at the level of uncertainty. It thus allows us to use the same cuts used in the experiment analysis, to get the number of events in each signal region. In top2(4) we fix $m_\chi(m_C)=1.1(1.5)m_{\wt t}$, and different ratios only change the kinematic distributions of final states slightly. More detail of the event simulation and check the validations of our implementing the experimental searches are discussed in Appendix~\ref{vali}.




\begin{table}[htb]
\begin{center}
\begin{tabular}{|c|c|c|c|c|} \hline
Number & Reference & Sensitive Channel \\
1  & \cite{ATLAS-CONF-2013-053}  & $\tilde{t} \to b l \tilde{\nu}$  \\
2  & \cite{ATLAS-CONF-2013-048}  & $\tilde{t} \to b l \tilde{\nu}$\\
3  & \cite{ATLAS-CONF-2013-037}  & $\tilde{t} \to t \nu \tilde{\nu}$\\
4  & \cite{ATLAS-CONF-2013-024}  & $\tilde{t} \to t \nu \tilde{\nu}$\\
5  & \cite{CMS-PAS-SUS-13-011}  & $\tilde{t} \to t \nu \tilde{\nu}$, $\tilde{t} \to b \tau \tilde{\nu}$\\
6  & \cite{ATLAS-CONF-2013-049}  & $C^\pm C^{\mp} \to 2 \times l \tilde{\nu}$\\
7  & \cite{ATLAS-CONF-2013-028}  & $\tilde{t} \to b \tau \tilde{\nu}$, $C^\pm C^{\mp} \to 2\times \tau \tilde{\nu}$\\
8  & \cite{ATLAS-CONF-2013-026}  & $C^{\pm}\chi\to \tau \nu \  2\times \tilde{\nu}$\\
9  & \cite{CMS-PAS-EXO-12-060}  & $C^{\pm}\chi\to l \nu \  2\times \tilde{\nu}$\\ \hline
\end{tabular}
\end{center}
\caption{The LHC searches used in this paper, where $l$ denotes either electron or muon. The number assigned for each search will be used in our plots to show the corresponding search which gives the strongest bound.}
\label{sLHC}
\end{table}

We briefly describe how to derive the bounds (see also Ref.~\cite{Cheng:2013fma}). In each signal region $S_i$ a variable $R_{vis}^i\equiv\frac{N^i_{\tiny{NP}}}{N_{limit}^i}$ is introduced. $N^i_{limit}$ and $N^i_{\tiny{NP}}$ respectively are the experimental upper limit and new physics contributed number of events. Then new physics with $N^i_{\tiny{NP}}$ will be excluded, as long as there exists any $R_{vis}^i>1$. The ATLAS collaborator explicitly gave $N_{limit}$, but CMS only presented the
observed number of events and expected number of background events
and the uncertainty in each signal region, so we have to derive the 95\% C.L. $N_{limit}$ from these data. This can be done via the standard Bayesian procedure:
\begin{align}
\frac{1}{\mathcal{N}} \int^{N_{limit}}_{0} \mathcal{L}(n_{obs}| N_{s}, N_b, \sigma_b) P(N_{s}) dN_s=0.95,
\end{align}
where $\mathcal{N}=\int^{\infty}_{0} \mathcal{L}(n_{obs}| N_{s}, N_b, \sigma_b) P(N_{s}) dN_s$, with a uniform prior probability $P(N_{s})$, is a normalisation factor. Taking into account the uncertainties of background and signal, we should take the following likelihood:
\begin{align}\label{ba}
\mathcal{L}(n_{obs}| N_{s}, N_b, \sigma_b) = \frac{1}{\sqrt{2 \pi \sigma_s^2} \sqrt{2 \pi \sigma_b^2}} \int^{5 \sigma_s}_{-5 \sigma_s} ds \int^{5 \sigma_b}_{-5 \sigma_b} db P(n_{obs};\mu) e^{\frac{db^2}{2 \sigma_b^2}} e^{\frac{ds^2}{2 \sigma_s^2}} ~.~\,
\end{align}
Practically, the probability function $P(n_{obs};\mu)$ takes  Possion distribution ${\mu^{n_{obs}} e^{- \mu}}/{n_{obs}!}$ for smaller $n_{obs} \leqslant 100$ while Gaussian ${e^{(n_{obs}-\mu)^2/2\mu}}/{\sqrt{2 \pi \mu}}$ for larger $n_{obs} >100$. Here the expectation value $\mu= n_s + ds + n_b + db$, and we assume the error $\sigma_s=0.01 \ n_s$. The derived $N_{limit}$ from Eq.~(\ref{ba}) will not change much as long as the signal uncertainty is within a few tens of percent, see Appendix~\ref{vali}.



\subsection{Bounds on chargino/sneutrino}
\label{C:bounds}
In this subsection we consider the bound on chargino/neutralino by decoupling stop. As have been discussed in Section~\ref{proddecay}, we have three kinds of decay topologies, depending on the patterns of production. Among them, we do not discuss the one from $\chi\chi$ production, which gives missing energy only. 
It can be probed only if an initial visible particle like a hard jet is emitted, but the current mono-jet plus $E_T^{\text{miss}}$ search~\cite{ATLAS-CONF-2013-068} is not able to impose a competitive bound, relative to the bounds from other decay topologies.

The only effective bound comes from $C^\pm C^\mp$ pair production followed by $C^{\pm} \to l \tilde{\nu}$. In what follows we will discuss two cases with $l=e/\mu$ and $\tau$, respectively. \begin{itemize}
   \item For the light lepton case, the decay topology gives rise to the signature of Opposite-Sign Dilepton (OSDL) plus large missing energy. The search for slepton pair production~\cite{ATLAS-CONF-2013-049} can probe this signature. In this search, the leptonic $m_{T_2}$ variable~\cite{Lester:1999tx,Barr:2003rg,Cheng:2008hk}
   \begin{equation}\label{MT2L}
m_{T_2} =\minmt_{q_T} \bigg[ \text{max}\big(m_T(p_T^{l1},q_T),m_T(p_T^{l2},p^{\text{miss}}_T-q_T)\big)\bigg],
\end{equation}
plays an important role to suppress the huge $t\bar t$ (and $WW$ as well) backgrounds, from which $m_{T_2}$ shows the sharp edge at $m_W$. While the edge of signal $m_{T_2}\simeq m_C$ is clearly larger than $m_W$, in particular for the heavier chargino. A large $E^{\text{miss}}_T$ may appear in the backgrounds, owing to the mis-measured momentum of jets or leptons. In order to reduce that, the quality $E^{\text{miss},\text{rel}}_T$ is introduced~\footnote{Its variants are frequently used by many groups, in any context where $E^{\text{miss}}_T$ is a crucial kinematic cut. A good case in point is the search of  $t\bar t$ plus $E^{\text{miss}}_T$ signature from stop pair production, as discussed later.}:
\begin{equation}
E^{\text{miss},\text{rel}}_T = \left\{
\begin{array}{rl}
 E^{\text{miss}}_T & \text{if}\,\, \Delta \phi_{l,j} \ge \pi/2 \\
 E^{\text{miss}}_T \times \sin \Delta \phi_{l,j} & \text{if}\,\, \Delta \phi_{l,j} < \pi/2 \\
\end{array}
\right. ~,~
\end{equation}
where $\Delta \phi_{l,j}$ is the azimuthal angle between the direction of $p_{T}^{\text{miss}}$ and its nearest lepton or jet. In addition, $Z-$veto is used to suppress the OSDL background from $Z$ decay.

The resulted constraints on charigno/sneutrino masses are similar to these on sleptons/neutralino masses in~\cite{ATLAS-CONF-2013-049}, but the exclusion limits is much higher by virtue of the larger cross section of chargino pair. The results are displayed in the upper panels of Fig.~\ref{emusv}. From it one can see that mass of (the Higgsino-like) chargino smaller than 550 GeV has been excluded, except that it degenerates with sneutrino, i.e., $m_{\tilde{H}}-m_{\tilde{\nu}} \lesssim 50$ GeV.
\begin{figure}[htb]
\begin{center}
\includegraphics[width=0.49\columnwidth]{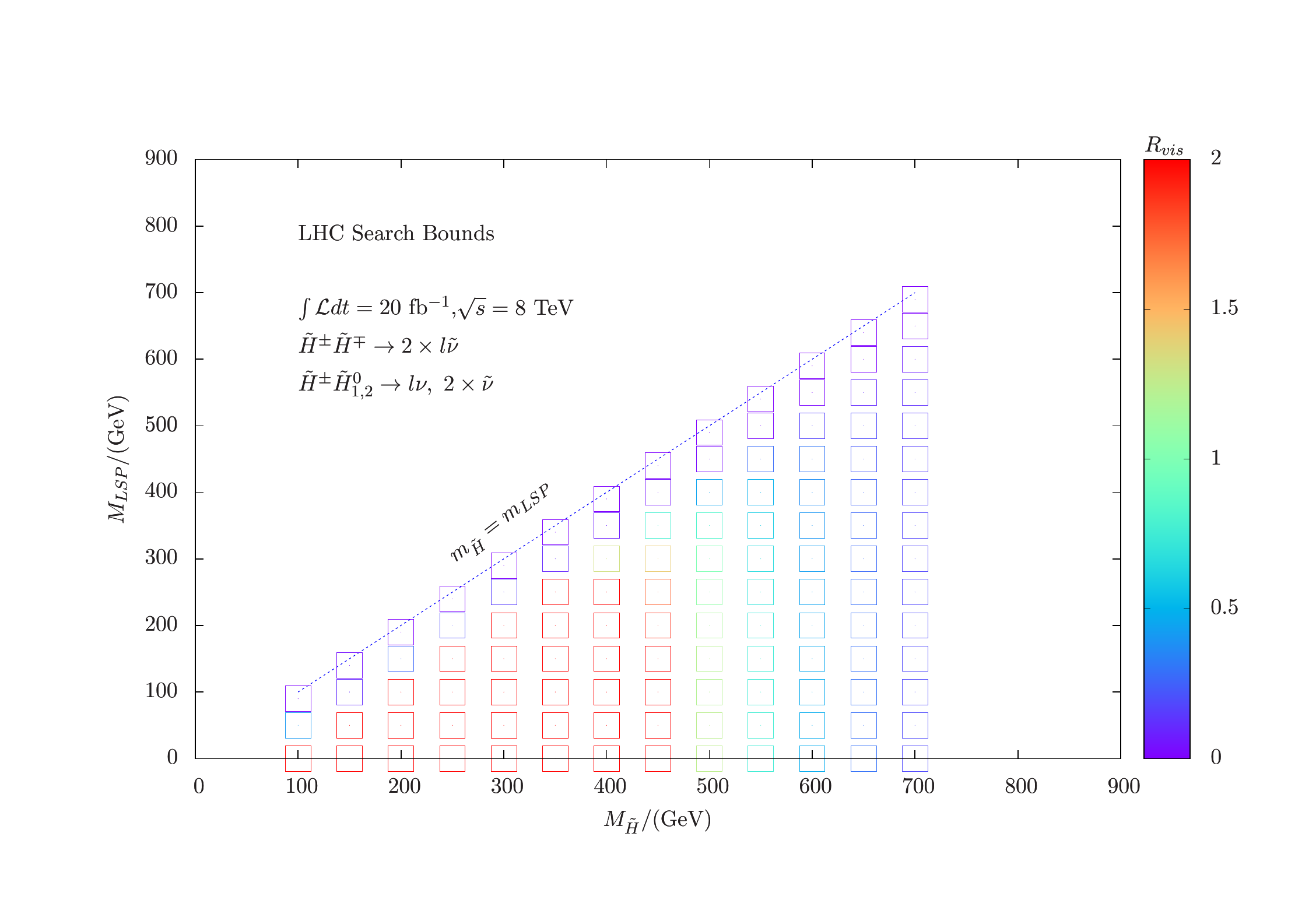}
\includegraphics[width=0.49\columnwidth]{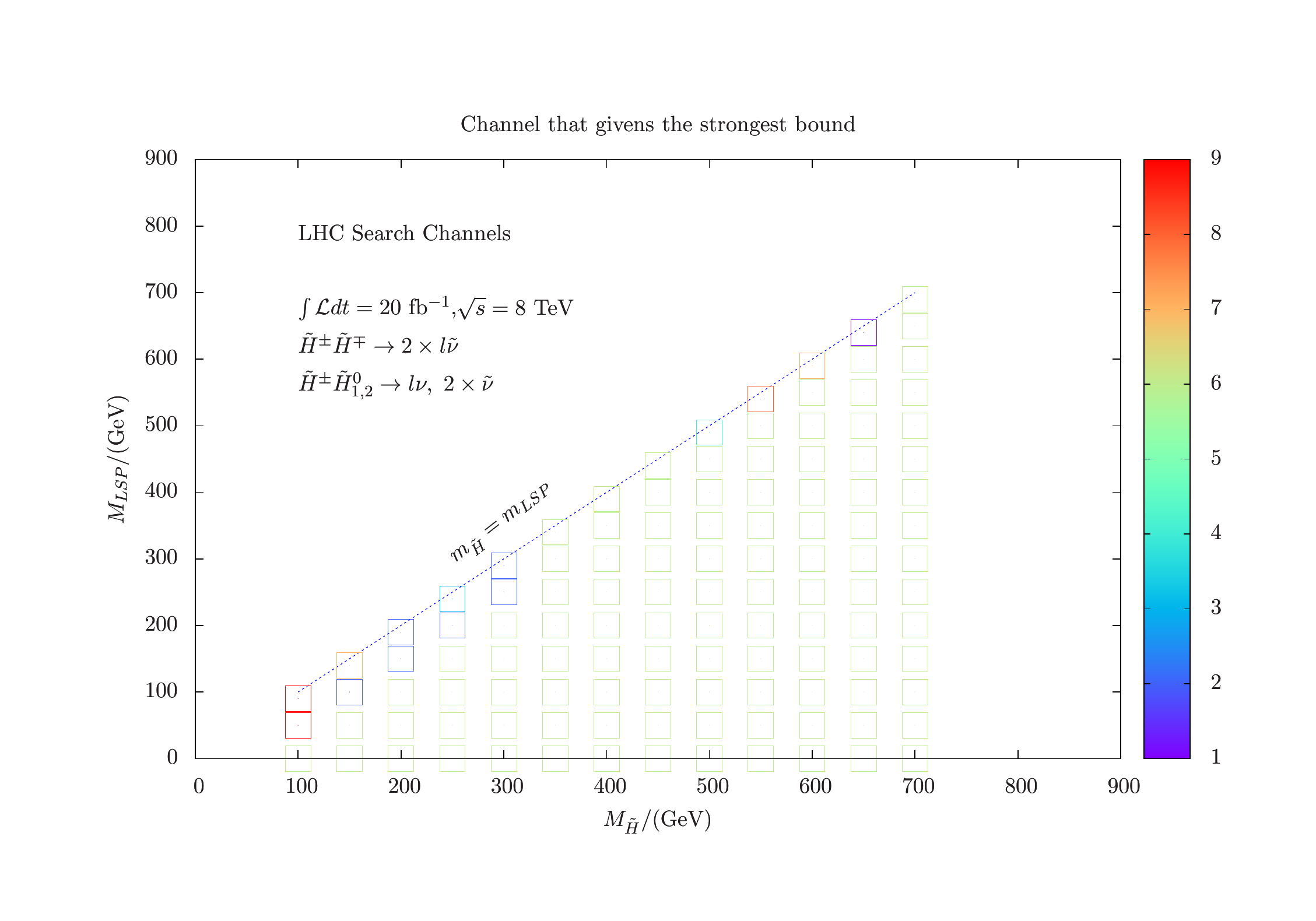}
\includegraphics[width=0.49\columnwidth]{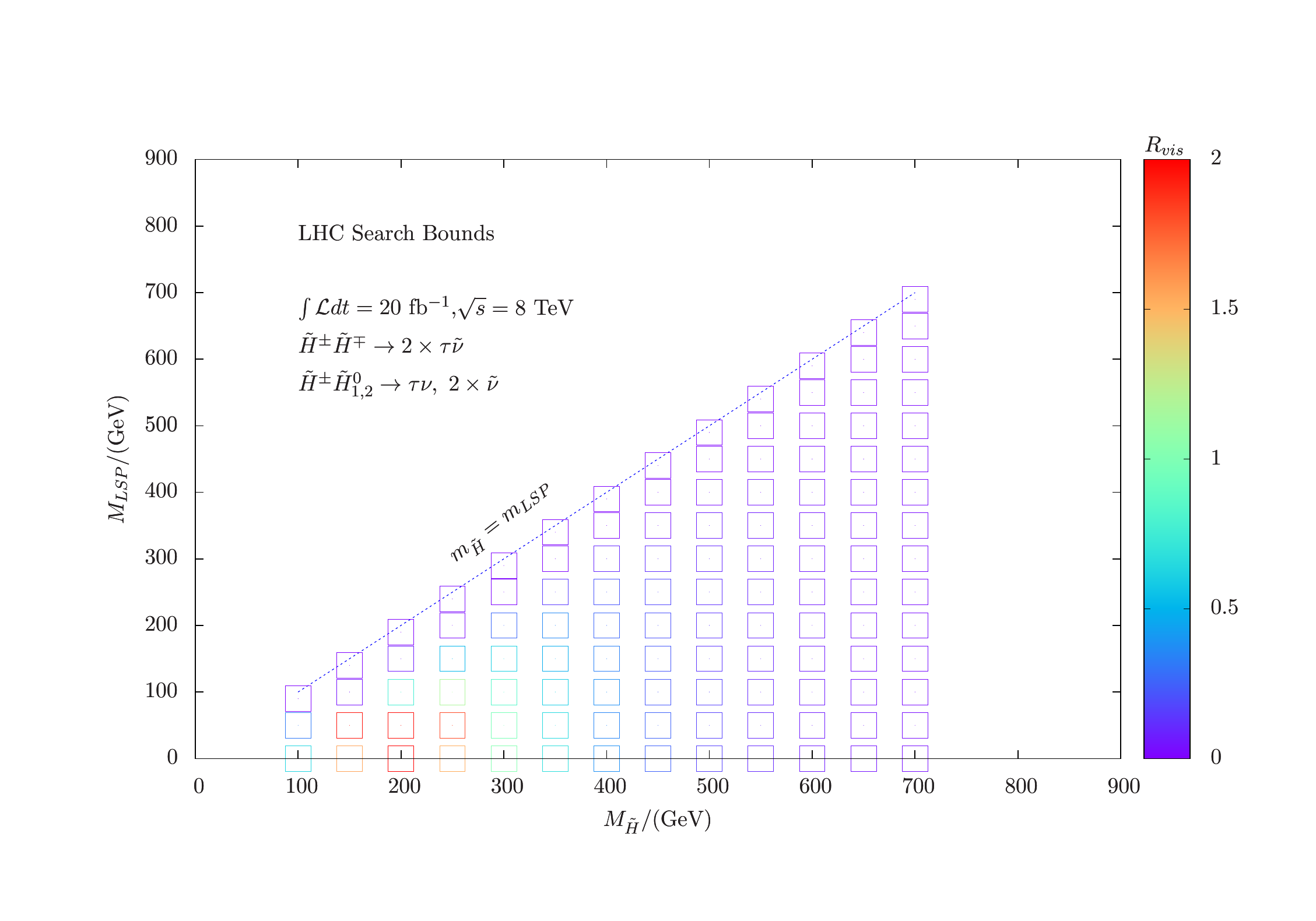}
\includegraphics[width=0.49\columnwidth]{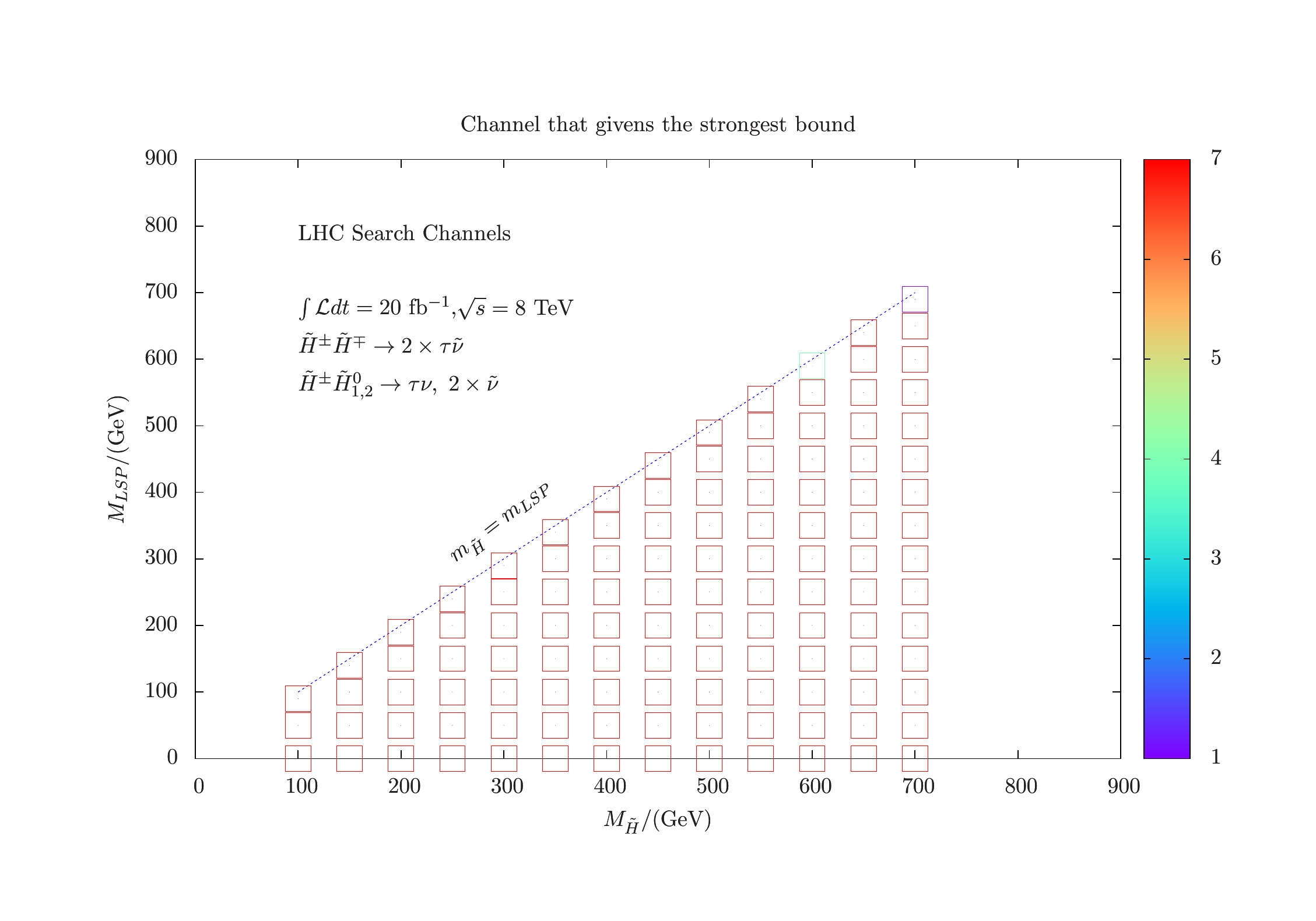}
\caption{Bounds on charigno/neutrlaino in the Higgsino-limit (Both $\tilde{H}^\pm\tilde{H}^\mp$ and $\tilde{H}^\pm\tilde{H}^0_{1,2}$ productions are included). Top (bottom) left: The maximal $R_{vis}$ (we have set any $R_{vis}$ larger than 2 to be 2) among all search channels for final states with $\ell$ ($\tau$); Top (bottom) right: The color coding of the channel which gives the most stringent bound, see Table~\ref{sLHC}. The same labels will be used throughout this paper. \label{emusv}}
\end{center}
\end{figure}

       \item In the case of $\tau$ final state, e.g., the sneutrino LSP is dominated by the third generation of sneutrino, the constraint becomes comparatively weaker. Among three decay modes of the $\tau-$pair, the one with both hadronically decaying $\tau$ acquires the most stringent bound in terms of Ref.~\cite{ATLAS-CONF-2013-028}. It searches the signature of two tau-jets plus large $E_T^{\text{miss}}$, using similar ways to the above case to suppress backgrounds.
         For the mode with both leptonically decaying $\tau$, the bound can be obtained as the $l=e/\mu$ case discussed before.  We show the final bounds in the bottom panels of Fig.~\ref{emusv}. The exclusion is significantly weaker than the previous case, and $m_C$ can only be excluded up to 350 GeV (for $m_{\tilde{\nu}}\lesssim 150$ GeV). The reason, asides from the branching ratio suppression, is mainly blamed to the low $\tau-$tagging efficiency and the softer $m_{T2}$.
                                      \end{itemize}

\begin{figure}[htb]
\begin{center}
\includegraphics[width=0.48\columnwidth]{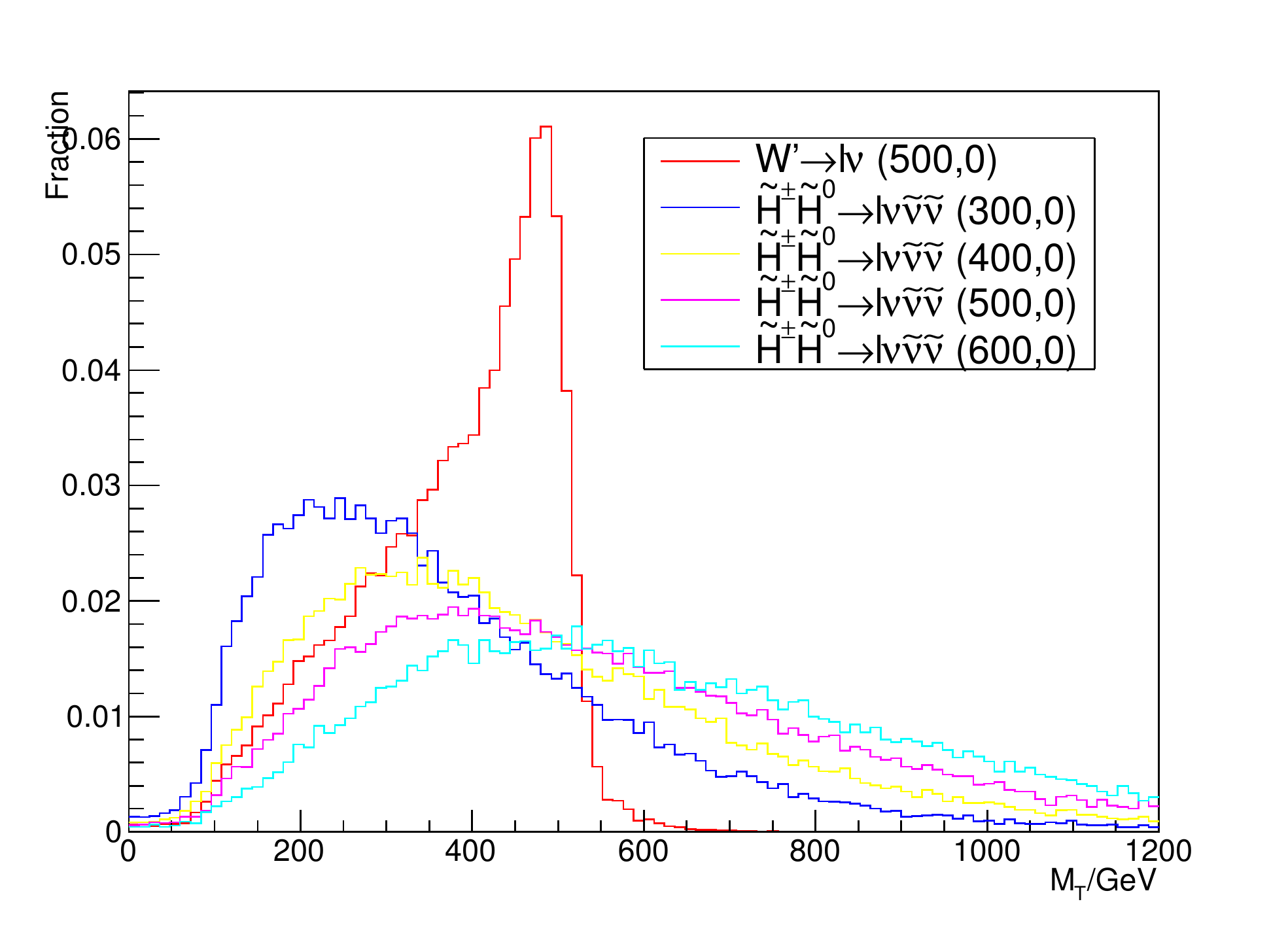}
\includegraphics[width=0.48\columnwidth]{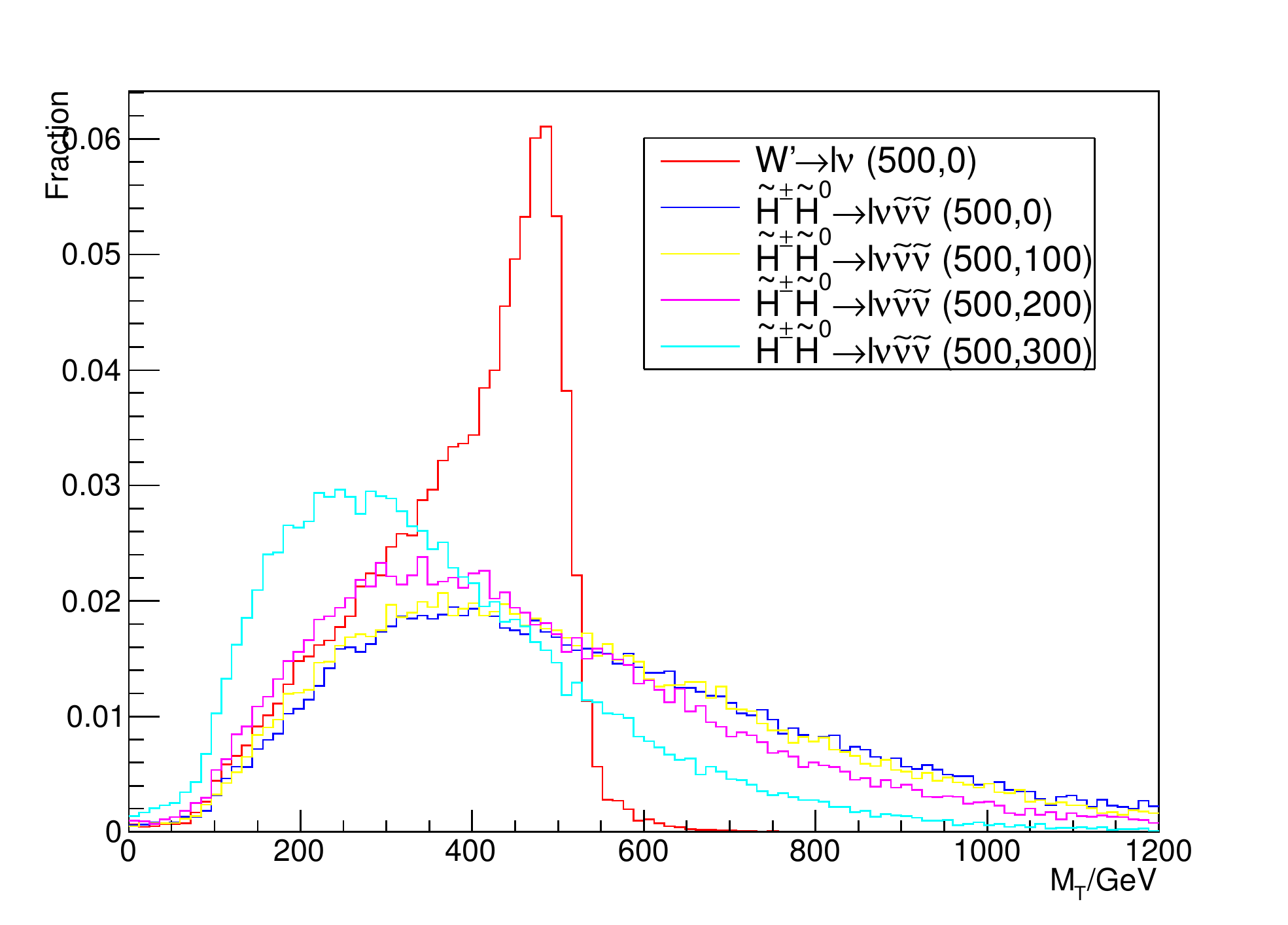}
\caption{$M_T$ distribution for $W'$ and Higgsino at detector level.
The corresponding masses are given in the Figure. \label{mtmono}}
\end{center}
 \end{figure}
 To end up this section, we would like to make a comment on the sensitivity to the $C/\chi$ associated production. From Fig.~\ref{xsec} one can see that it has an even larger cross section than the chargino pair production, given $C/\chi$ in the Higgsino/wino-limit. Consider first the mono-lepton ($e/\mu$) plus $E_T^{\text{miss}}$ signature, to which the only known relevant search by the CMS Collaboration~\cite{CMS-PAS-SUS-13-011} is not sensitive. The reason is simple: The CMS search heavily relies on the $M_T$ variable, which however has only been optimized for the heavy $W'$ ($\sim 1$ TeV) search. Consequently, it fails in imposing bounds on the relatively light $C/\chi$, see Fig.~\ref{mtmono}. But $M_T$ is indeed a good variable to probe new physics, so hopefully we can improve the low $M_T$ region, by using additional cuts, to enhance sensitivity to this channel. The mono-$\tau$ channel, which gives rise to the mono-jet plus $E_T^{\text{miss}}$ signature, deserves special attention. Now, exclusion on new physics cross section from mono-jet search is merely about 100 fb~\cite{ATLAS-CONF-2013-068}. But jet flavor tagging may help much. Using the extrapolation from charm-tagging in the mono-jet search~\cite{ATLAS-CONF-2013-068}, the sensitivity is enhanced by about two magnitudes of order and reaches $\sigma>0.7$ pb (roughly corresponding to a bound $m_C>600$ GeV in the Higgsino limit).

\subsection{Bounds on stop/sneutrino}

Bounds on stop are complicated due to its rather rich decay topologies, top1-4. But top1 is identical to the ordinary SUSY case with stop pair production followed by $\wt t\ra t\chi$, so in what follows we focus on top2-4.

\subsubsection{Bounds from top2: $\wt t\ra t \wt \nu\nu$}

The final states of stop pair production in this channel contain a pair of top quarks plus missing energy, just the same as those of
$\wt t\wt t^*\ra t\bar t+2\chi$ in the conventional SUSY scenario. As a matter of fact, the latter is expected to be the bulk signature in the supersymmetric models like MSSM and careful searches for it have been performed~\cite{ATLAS-CONF-2013-024,ATLAS-CONF-2013-037}. Hence they can be used to well probe stop/sneutrino in our paper.


The most sensitive searching channel is different in different stop/sneutrino mass regions, depending on the efficiency of reconstructing the top quark. When their mass difference is relatively large ($\gtrsim 300$ GeV), top quark from the heavy stop decay can be reconstructed rather effectively in a small cone and then search for the hadronic decaying stop pair, aided by a large $E_T^{\text{miss}}$, is able to give the strongest bound~\cite{ATLAS-CONF-2013-024}. When the mass difference becomes smaller, top reconstruction becomes worse and hence the semi-leptonic stop pair, which gives one isolated hard lepton, will instead impose the strongest bound~\cite{ATLAS-CONF-2013-037}. As the mass difference further decreases, namely it enters into the degenerate region, the signatures turn out to be almost indistinguishable from the $t\bar t$ background. Thus here we fail to make bounds on the stop/sneturino masses and should turn to other strategies~\cite{Ajaib:2011hs,Yu:2012kj,Han:2013usa}. But this topic is beyond the scope of this work.
\begin{figure}[htb]
\begin{center}
\includegraphics[width=0.48\columnwidth]{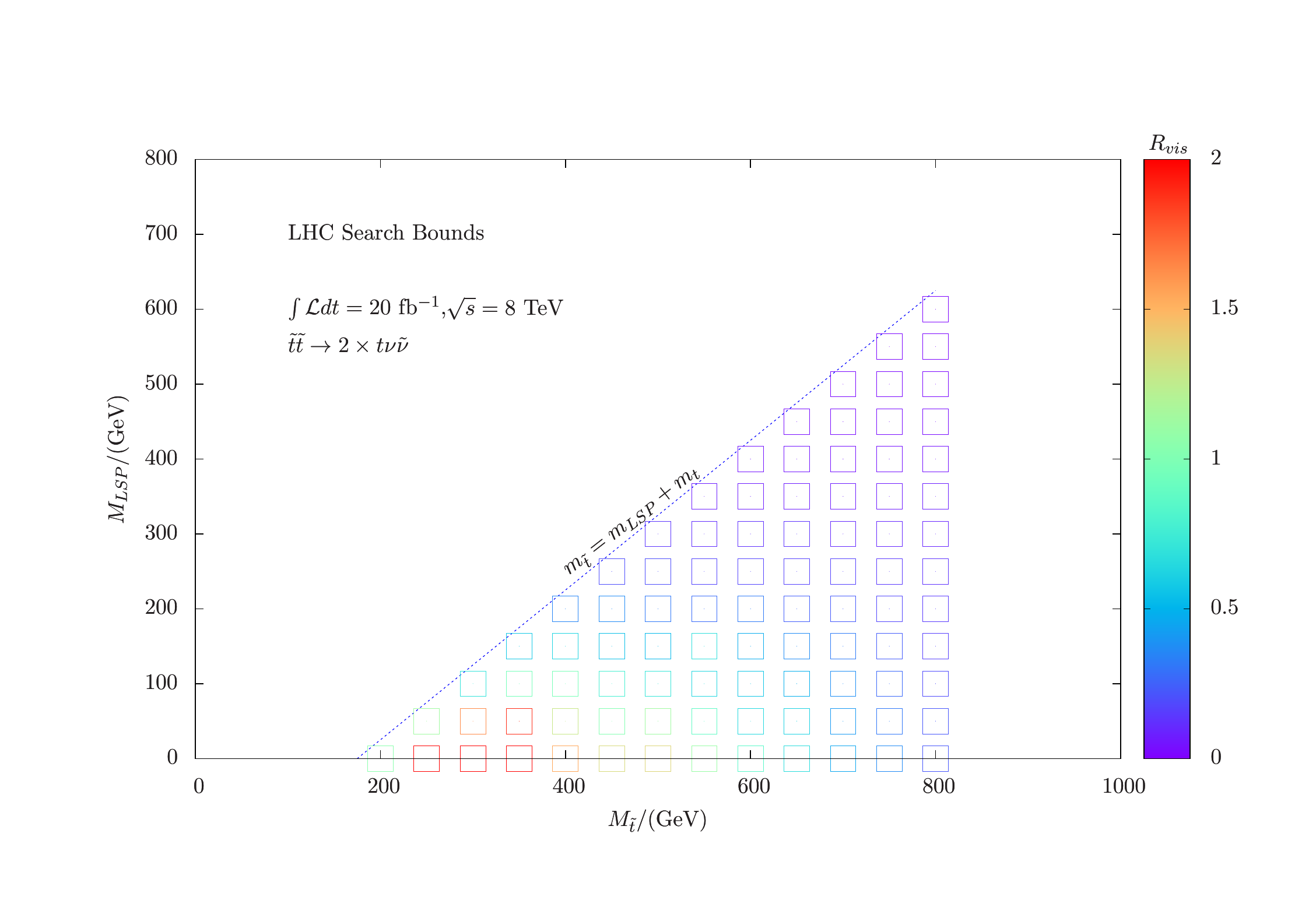}
\includegraphics[width=0.48\columnwidth]{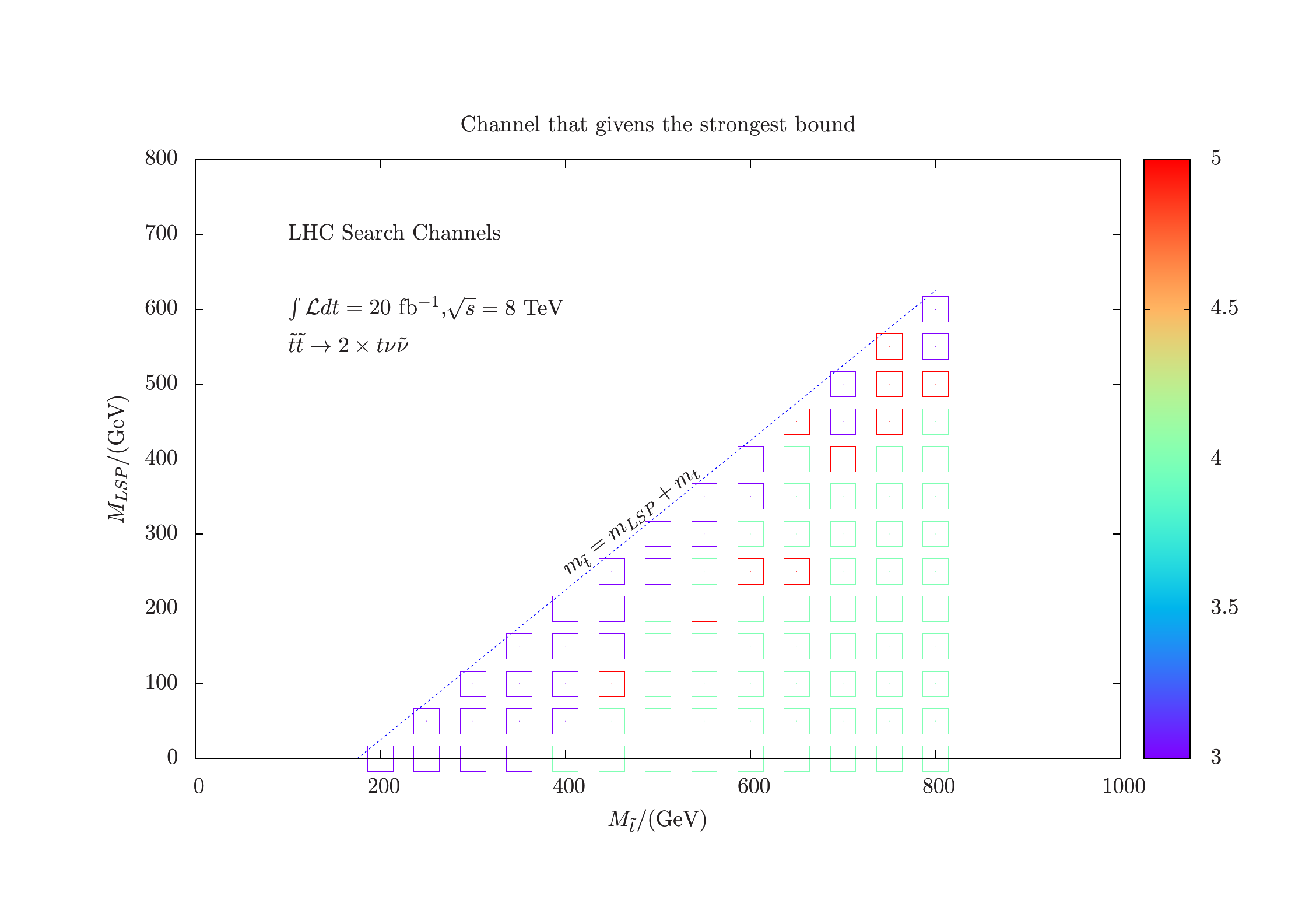}
\caption{Bounds on stop/sneutrino masses from the $\tilde{t} \to t \nu \tilde{\nu}$ channel. Color scheme is the same as Fig.~\ref{emusv}.
\label{Atn1_us}}
\end{center}
\end{figure}

The resulted bounds on stop and sneutrino masses are shown in Fig.~\ref{Atn1_us}. It is seen that the exclusion limits have reached $m_{\tilde{t}}\lesssim 600$ GeV and $m_{\tilde{\nu}}\lesssim 100$ GeV, but it is mildly weaker than the ordinary case $\tilde{t} \to t {\chi}$, which has reached the region $m_{\tilde{t}}\lesssim 700$ GeV and $m_{\tilde{\nu}}\lesssim 250$ GeV. This is not surprising. In top2 stops are three-body decaying,
which renders the kinematic distributions of the finals states more dispersed than these of the ordinary two-body decay. Especially, the missing energy is substantially softened. These features lead to the decreased sensitivity to the signature when we use data of searching stop pair with $\wt t\ra t\chi$ to probe top2.

Actually, one can extend the current exclusion limit on the heavier stop ($\gtrsim 600$ GeV), by using HEPTopTagger~\cite{Plehn:2010st} instead of the algorithm taken by experimentalists to tag the hadronic decaying tops.Even requiring the reconstructed top mass within a narrow window,$[150,200]$ GeV, HEPTopTagger can tag the top-jet with $p_T>200$ GeV efficiently ($\sim37\%$). Thus it may help to improve our search for three-body decaying stop which produces relatively soft top. We have checked this for a mildly heavy stop with mass $600$ GeV (and $m_{\tilde{\nu}}=100$ GeV), to find that the top-tagging efficiency still reaches 20.77\% before the $p_T$ cut.

\subsubsection{Bounds from top4: $\tilde{t} \ra b\,l\, \tilde{\nu}$}

We postpone the more complicated case, top3, to the last subsubsection and here we focus on top4 first. As before, we respectively discuss two cases $l=e/\mu$ and $l=\tau$.
\begin{itemize}
  \item For $l=e/\mu$, the final states contain two hard $b$-jets, two hard leptons with opposite sign (i.e., OSDL) and large missing energy. We can explore such a signature through the search for stop pair production followed by $\tilde{t} \to b C$ then $C \to W^* \chi$, which is another conventional bulk signature for stop~\cite{ATLAS-CONF-2013-053}. This search is somewhat like the search of slepton pair production discussed previously, where the leptonic $m_{T2}(\vec{p}_T^{l1},\vec{p}_T^{l2},p^{\text{miss}}_T)$ is an useful discriminator. And the presence of two hard $b$-jets further enhances the sensitivity to the signature. Note that the experimental cut on $m_{T2}$ is relatively low, which benefits our case, where stop three-body decay leads to a softer $m_{T2}$.

      The bounds are displayed in the top panels of Fig.~\ref{Abc1_us}. The bounds are fairly stringent, e.g., the stop mass have been excluded up to about 900 GeV, leaving only the degenerate region where the mass difference between $\tilde{t}$ and the LSP is smaller than 50 GeV. These exclusions are even more stringent than the exclusions for stop/neutralino in Ref.~\cite{ATLAS-CONF-2013-053}, since there OSDL comes from the leptonic $W$ pair, suppressed by branching ratio. On top of that, here the leptons are directly produced from stop decay, so they can be quite hard.
\begin{figure}[htb]
\begin{center}
\includegraphics[width=0.48\columnwidth]{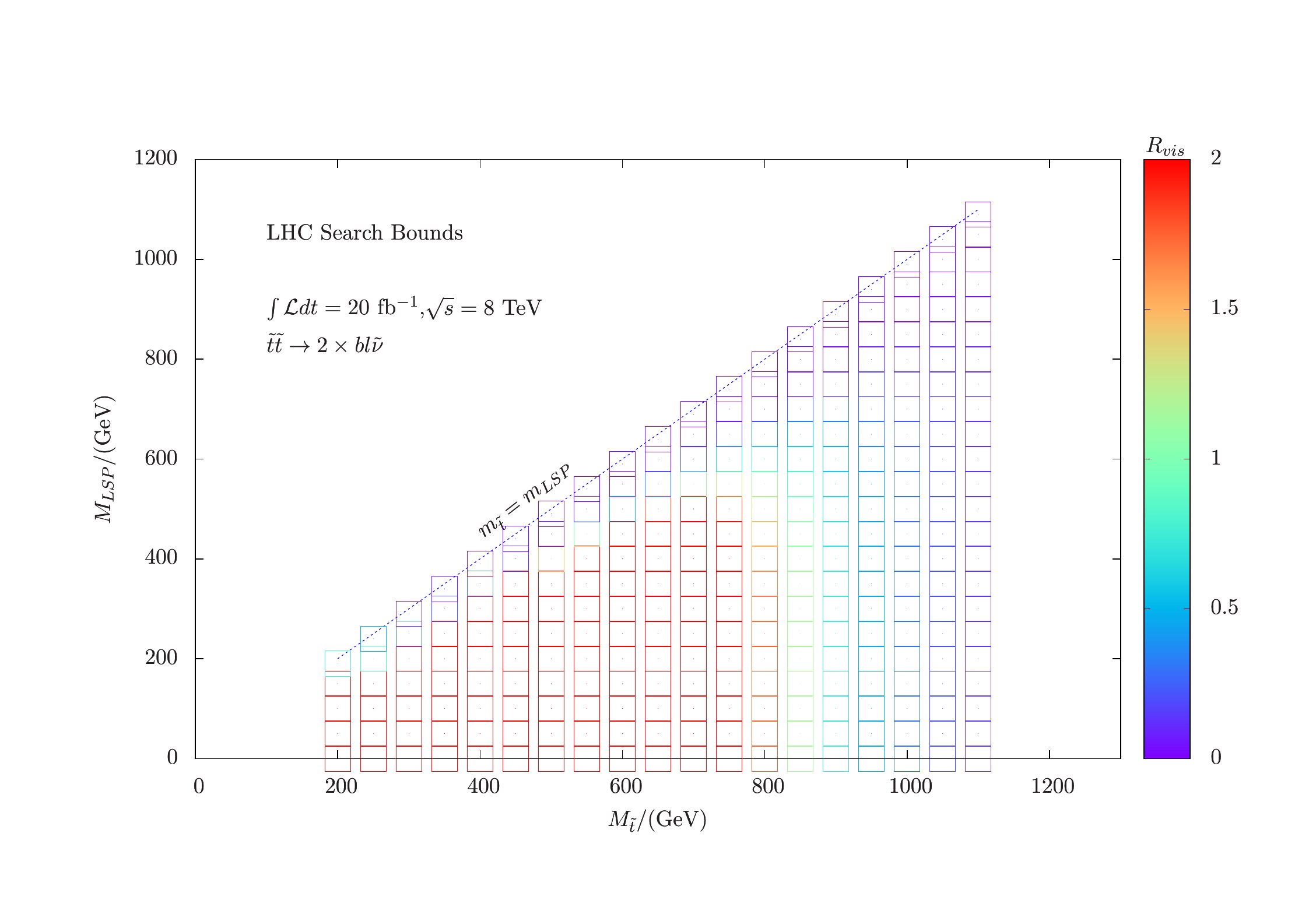}
\includegraphics[width=0.48\columnwidth]{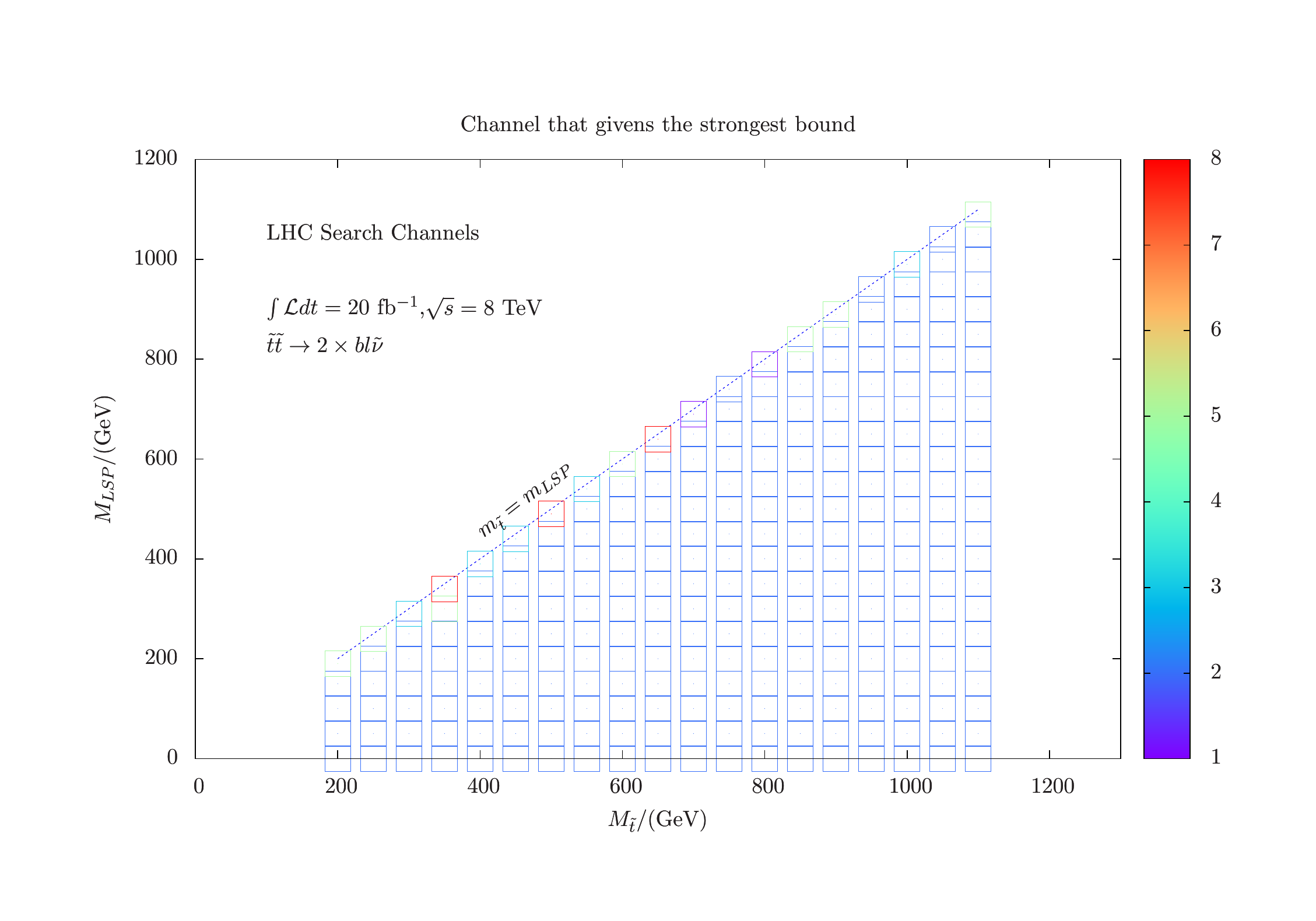}
\includegraphics[width=0.48\columnwidth]{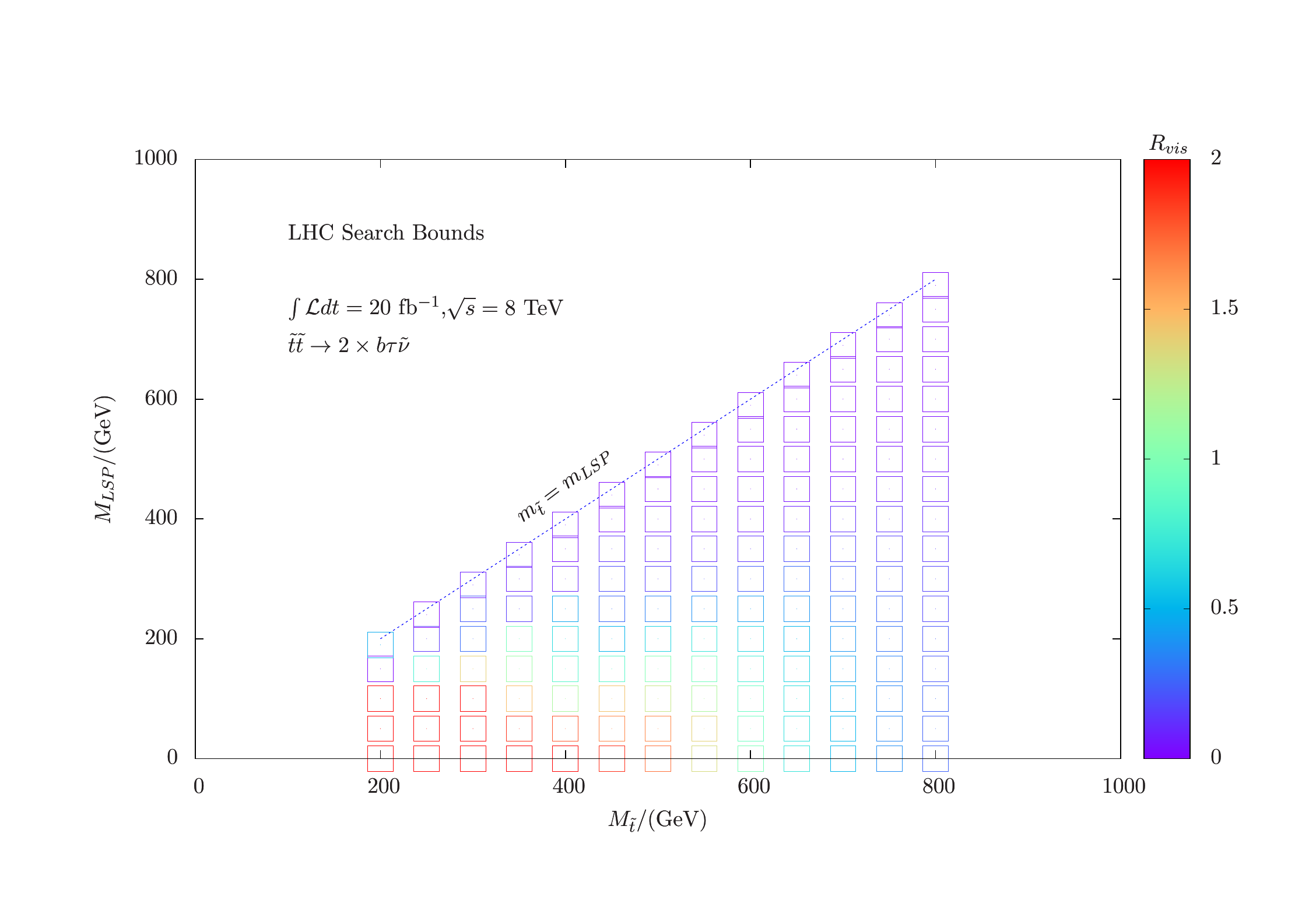}
\includegraphics[width=0.48\columnwidth]{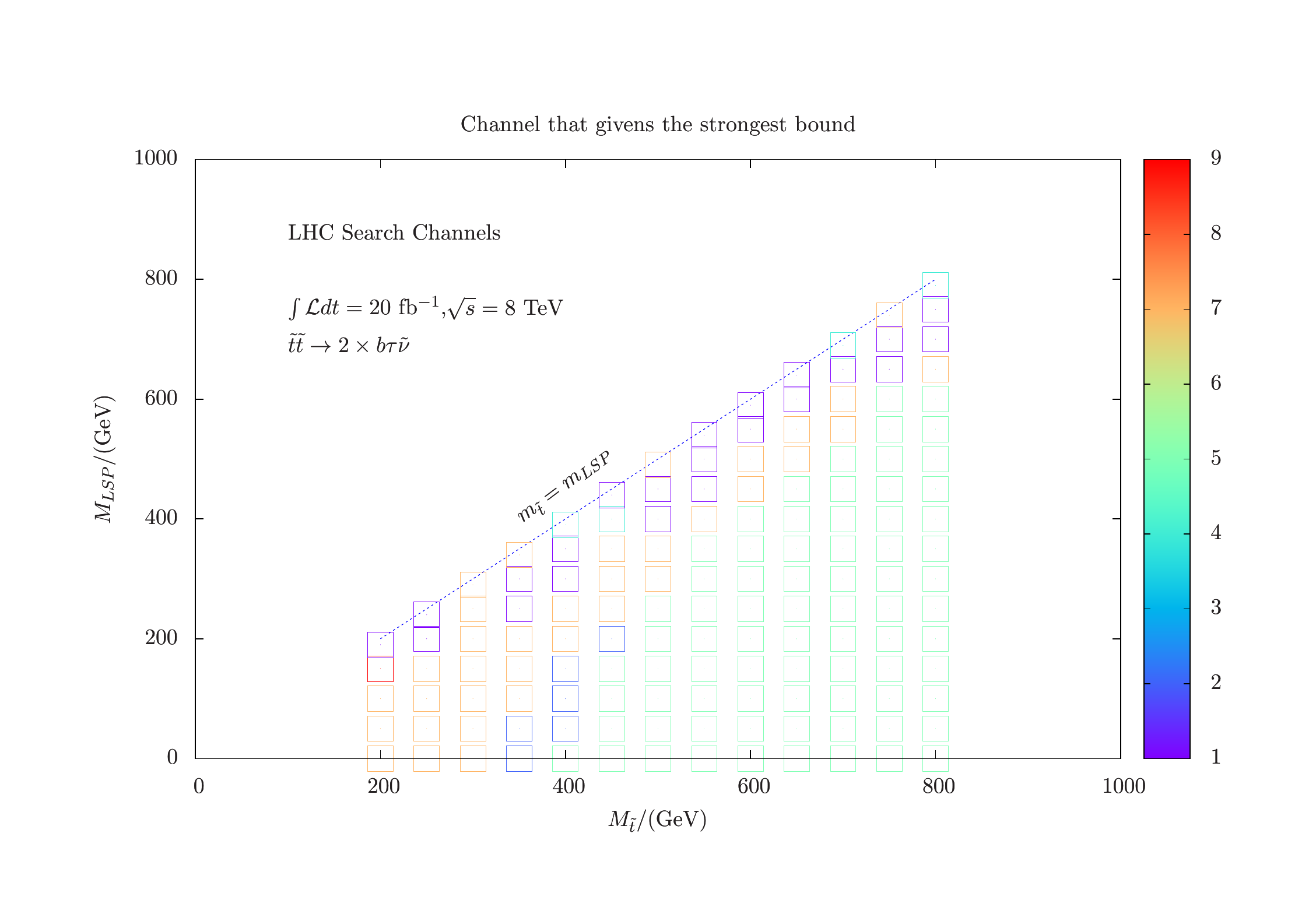}
\caption{Bounds on stop/sneutrino masses. Top panels: from the $\tilde{t} \to b l \tilde{\nu}$ channel; Bottom panels: from the $\tilde{t} \to b \tau \tilde{\nu}$ channel. Color scheme is the same as Fig.~\ref{emusv}.
\label{Abc1_us}}
\end{center}
\end{figure}
  \item For $l=\tau$,  searches depend on the decay of $\tau$. In the case of two hadronically decaying taus, the final states contain two tagged $b$- and $\tau$-jets. Despite of
      no direct searches for such a signature, we can still probe them indirectly by means of the searches which tag two hadronically decaying taus or two $b$-jets. Similarly we can have sensitivity to the semi-leptonic decaying $\tau$ case. As for the two leptonic decaying taus case, the resulted signature actually is the same as the one discussed above. But here its cross section is suppressed by the squared $\tau$  leptonic branching ratio ${\rm {Br}}^2$($\tau\ra \ell \nu\nu)\simeq0.1$.

The current bounds on this channel are shown in the bottom panels of Fig.~\ref{Abc1_us}. The stop mass is excluded up to about 600 GeV for $m_{\wt\nu}$ below 150 GeV. The bound on the heavier stop region is set by the CMS search~\cite{CMS-PAS-SUS-13-011} for signature of two $b$-jets, one lepton and $E_T^{\text{miss}}$. While at the lighter stop region with $m_{\tilde{t}}<350$ GeV and the degenerate region as well, the strongest bound comes from the search for final state with two tau-jets plus $E_T^{\text{miss}}$~\cite{ATLAS-CONF-2013-053} or two $b$-jets plus $E_T^{\text{miss}}$~\cite{ATLAS-CONF-2013-028}. And here the sneutrino mass has been excluded up to 200 GeV.

The bounds on this case will be further improved if tau is tagged out. Naively, the tau-richness in the final state is able to  suppress the background by an order of magnitude, since tau in the background mainly comes from $W/Z$ boson decays, suppressed by the small branching ratios. Then, we may be able to push the bound on stop much above 600 GeV in the semi-leptonic tau channel.
\end{itemize}

\subsubsection{Bounds from top3: $\tilde{t} \to b \chi\ra  b\,l\, \tilde{\nu}$}

Top3 and top4 share the same final states, but bounds on the former involve an extra parameter, mass of the chargino $m_C$. As a result, making a comprehensive bound becomes much more complicated, and we have to take several typical values of $m_C$ and then investigate the corresponding bounds on stop/sneutrino. Concretely, three typical cases are considered (we focus on $l=e/\mu$ case):
\begin{itemize}
\item The chargino mass is close to the sneutrino mass. In this case the lepton from chargino decay is too soft to be detected and then chargino behaves as a missing particle at the collider. The signature is two hard $b$-jets plus large $E_T^{\text{miss}}$, just the same as that of the ordinary channel $\tilde{t}\to bC^\pm,\,C^\pm \to W^* \chi$ with $m_C\simeq m_\chi$. From Ref.~\cite{ATLAS-CONF-2013-053} the stop mass has been excluded up to 600 GeV.
\item Instead, the chargino mass is close to the stop mass, rendering the soft $b$-jets invisible. Evidently, this case is reduced to chargino pair production with a rescaled production cross section. Therefore, in the light of the bounds on charigno made in Section.~\ref{C:bounds} and the production cross sections in Fig.~\ref{xsec}, we estimate the bound on the stop mass,  $\sim$900 GeV.
\item Generically, the chargino mass is neither close to the sneutrino nor the stop mass. In this case the signature is identical to that of top4 and bounds again can be derived in the light of Ref.~\cite{ATLAS-CONF-2013-053}. We consider three representative examples for charigno mass: $m_{\tilde{\chi}^\pm}=0.8\times m_{\tilde{t}}$, $m_{\tilde{\chi}^\pm}=0.5\times m_{\tilde{t}}$ and $m_{\tilde{\chi}^\pm}=0.2\times m_{\tilde{t}}$. In each example, the strongest bound on stop mass is reached for a very light sneutrino. Concretely, it can be excluded up to about 850 GeV, 850 GeV and 600 GeV, respectively.  The comparatively weaker bound on the third example is due to the smallness of $m_{T2}(\vec{p}_T^{l1},\vec{p}_T^{l2},p^{\text{miss}}_T)$, which reflects the light chargino mass.
\end{itemize}

\subsubsection{$m_{T2}$ distinguishes the three-body from two-body decays}

For a given process, signatures of tree-body and two-body decays are the same with each other, e.g., stop pair production in top1 and top3, top2 and top4. Thus, (after the discovery of event excess) it is of interest to explore a kinematic  variable to tell the difference. Specific to our article, $m_{T2}$ is a good candidate. 

\begin{figure}[htb]
\begin{center}
\includegraphics[width=0.48\columnwidth]{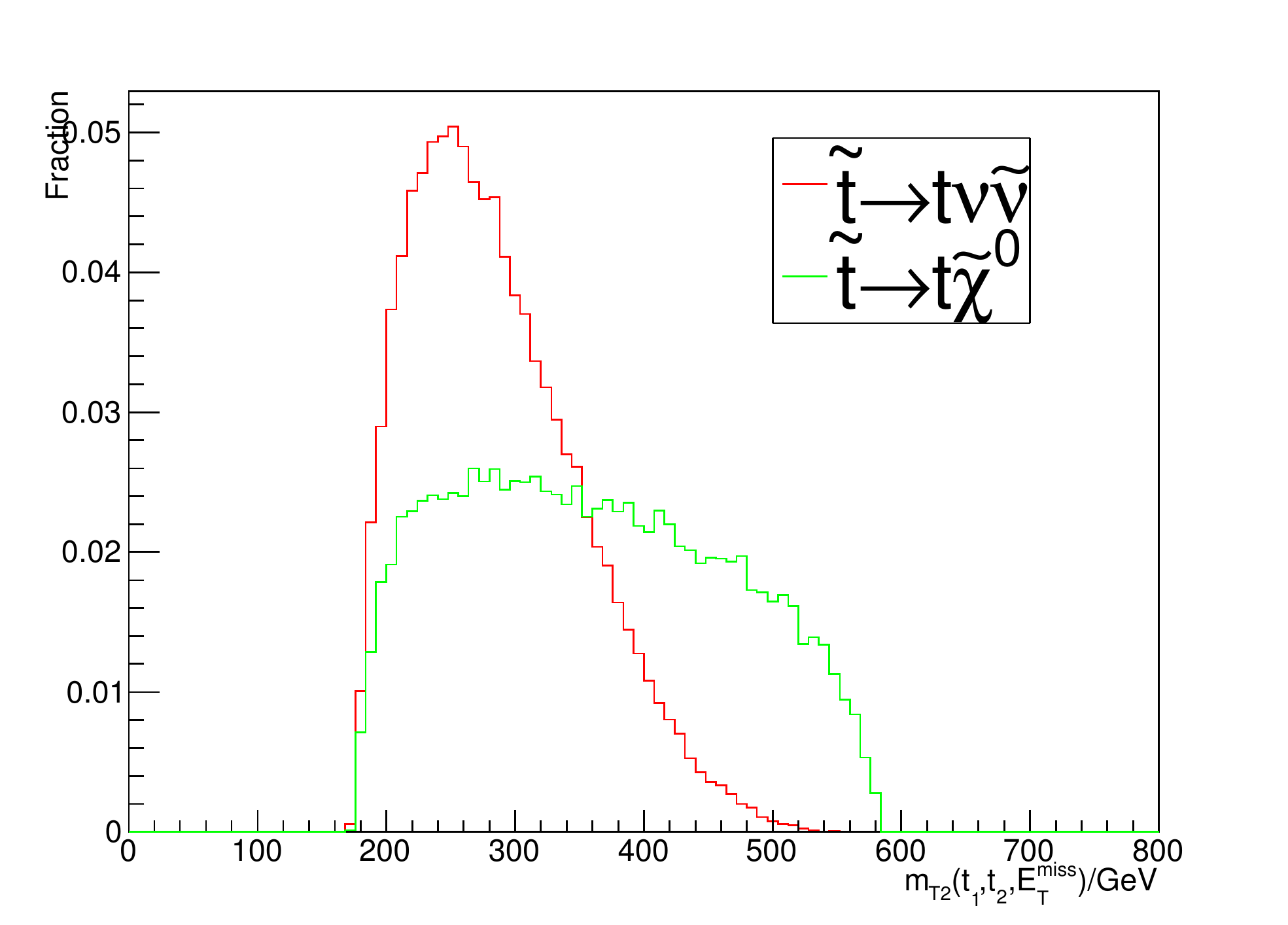}
\includegraphics[width=0.48\columnwidth]{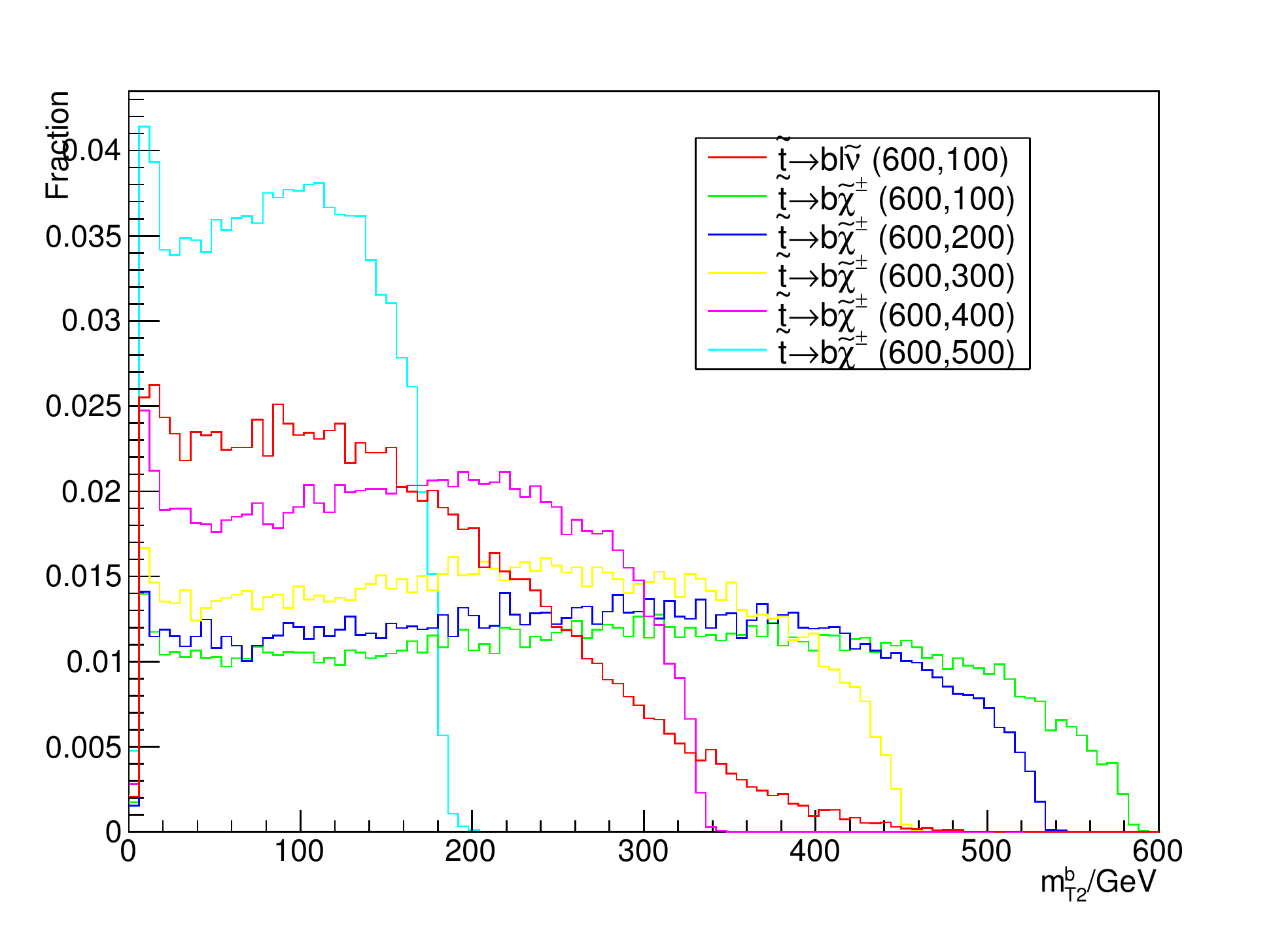}
\caption{ Comparisons of the $m_{T2}$ distributions for two- and tree-body stop decays at parton level. Left: $m_{T2}(t_1,t_2,p^{\text{miss}}_T)$ for decays mediated by neutralino. We take $m_{\tilde{t}}=600$ GeV and $m_{\text{LSP}}=100$ GeV. Right: $m_{T2}^b$ for decays mediated by chargino. We take $m_{\tilde{t}}=600$ GeV and vary the charigno mass.
\label{mt2}}
\end{center}
\end{figure}

First consider top1 (equivalent to $\wt t\ra t\chi$) and top3. With the reconstructed top quarks, we can construct $m_{T2}(t_1,t_2,p^{\text{miss}}_T)$. And we plot the distributions for top1 and top3 in the left panel of Fig.~\ref{mt2}. Their shapes are clearly different: The curve for top1, namely for two-body decay, is much flatter and drops suddenly near $m_{\wt t}$; By contrast, the curve for top3, the three-body decay case, drops much earlier and faster. Alternatively, one can tell the difference by observing the property of curve at the tail. In two- and three-body decay they are convex and concave, respectively. Similar conclusion applies to top2 and top4, for which we use the modified $m_{T2}$, i.e. $m^b_{T2}$~\cite{Bai:2012gs}:
\begin{equation}
m^b_{T2} =\text{min} \left\{  \bigcup_{\vec{p}_1^T+\vec{p}^T_2=\vec{p}^{\text{miss}}_T+\vec{p}^T_{l1}+\vec{p}^t_{l2}} \text{max} [M_T(\vec{p}_{b_1},\vec{p}^T_1), M_T(\vec{p}_{b_2}, \vec{p}^T_2)] \right\} ~.~
\end{equation}
where the leptons are also identified as missing energy and the bottom quarks are the only visible particles.

\section{Conclusions and Discussion}

Due to a successful WIMP dark matter candidate, the lightest neutralino LSP is assumed to be the termination of sparticle cascade decays. It is the premise of most SUSY searches. But sneutrino LSP can also be a good thermal DM candidate in many supersymmetric SMs with low-scale seesaws. Thus it is well motivated to explore SUSY with sneutrino LSP. We construct a simplified model to describe their collider phenomenologies. Asides from the sneutrino, the model  contains a stop, charigno and neutraino which can promptly decay into sneutrino at the LHC.

As a result of the leptonic nature of the sneutrino, leptons, associated with sneutrino, appear in the decay topologies of sparticles.  This means that they may leave clear tracks at the collider. In fact, we find that the current SUSY searches at the LHC have already put strong bounds on them:
\begin{itemize}
\item  For $C\ra \ell \wt\nu$, the bound on charino/neutrino masses in the Higgsino-limit reaches 550 GeV, leaving the degenerate region with $m_{\tilde{H}}-m_{\wt \nu}<50$ GeV untouched. But for $C\ra \tau \wt\nu$ the bound is substantially weakened, only excluding $m_C\lesssim$ 350 GeV and $m_{\wt\nu}\lesssim$ 150 GeV. This is owing to both the inefficient $\tau-$tagging and the absence of specific searches.
  \item Mass of the NLSP stop has been excluded up to about 600 GeV for $\wt t\ra t\nu\wt\nu$ and up to about 900 GeV for $\wt t\ra b \ell \wt\nu$. In the former case, $m_{\wt \nu}$ is excluded up to about 100 GeV while in the latter case only sneutrno in the degenerate region survives. For $\wt t\ra b \tau \wt\nu$, compared to the $\tilde{t} \to b l \tilde{\nu}$ channel, the current searches have much weaker sensitivities not only for the heavier stop above $600$ GeV but also for the heavier sneutrino with mass above 200 GeV.
     \item If stop is not the NLSP, it will decay into neutralino or  chargino  first. The former case is just the same as the usual neutralino LSP scenario. The latter case can be divided into three categories, depending on the chargino mass. The stop mass have been excluded up to about 600 GeV/900 GeV and 850 GeV for chargino mass close to the sneutrino/stop mass and close to neither sneutrino nor stop mass.
\end{itemize}
We also briefly discuss possible optimizations for their searches without turning to the future LHC running.

\section{Acknowledgment}

This research was supported in part by the China Postdoctoral Science Foundation (No.
2012M521136) (ZK), and Natural Science Foundation of China under
grant numbers 10821504, 11075194, 11135003, and 11275246 (TL).

\appendix
\section{The Stop and Higgisno Decays}

\label{adecay}

In this appendix, we give the general expressions for the stop and chargino/neutralino decay widths. The Feymann diagrams are shown in Fig.~\ref{procs}. For the stop three-body decaying into sneutrino, we have
{\small\begin{align}\label{gammastop}
\Gamma=&\frac{1}{128\pi}\frac{M_{\widetilde{t}}^5}{M_{\widetilde{\chi}^4}}\bigg( g_L^q g_R^q (g_L^l)^2+g_L^q g_R^q(g_R^l)^2\bigg)\widehat{f_{3}}\cr
        +&\frac{1}{256\pi}\frac{M_{\widetilde{t}}^5}{M_{\widetilde{\chi}^4}}\bigg((g_L^q)^2(g_R^l)^2+(g_R^q)^2(g_L^l)^2\bigg)\widehat{f_{2}} +\frac{1}{256\pi}\frac{M_{\widetilde{t}}^5}{M_{\widetilde{\chi}^4}}\bigg((g_L^q)^2(g_L^l)^2+(g_R^q)^2(g_R^l)^2\bigg)\widehat{f_{1}} ~,~\,
\end{align}}where $\widehat{f_{i}}$ is the three-body final state phase space integral, given by
{\small\begin{align}
\widehat{f_1}=&\frac{m_{\widetilde{\chi}}^6}{m_{\widetilde{t}}^7}\int^{(m_{\widetilde{t}}-m_q)^2}_{m_{\widetilde{\nu}}^2} d P_{\widetilde{\chi}}^2 \frac{ \sqrt{\bigg(\frac{m_{\widetilde{t}}^2+P_{\widetilde{\chi}}^2-m_q^2}{2m_{\widetilde{t}}}\bigg)^2-P_{\widetilde{\chi}}^2}(P_{\widetilde{\chi}}^2-m_{\widetilde{\nu}}^2 )^2(m_{\widetilde{t}}^2-m_q^2-P_{\widetilde{\chi}}^2)}{P_{\widetilde{\chi}}^4(P_{\widetilde{\chi}}^2-m_{\widetilde{\chi}}^2)^2} ~,~\cr
\widehat{f_2}=&\frac{m_{\widetilde{\chi}}^4}{m_{\widetilde{t}}^5}\int^{(m_{\widetilde{t}}-m_q)^2}_{m_{\widetilde{\nu}}^2} d P_{\widetilde{\chi}}^2 \frac{ \sqrt{\bigg(\frac{m_{\widetilde{t}}^2+P_{\widetilde{\chi}}^2-m_q^2}{2m_{\widetilde{t}}}\bigg)^2-P_{\widetilde{\chi}}^2}(P_{\widetilde{\chi}}^2-m_{\widetilde{\nu}}^2 )^2(m_{\widetilde{t}}^2-m_q^2-P_{\widetilde{\chi}}^2)}{P_{\widetilde{\chi}}^2(P_{\widetilde{\chi}}^2-m_{\widetilde{\chi}}^2)^2} ~,~\cr
\widehat{f_3}=&\frac{m_{\widetilde{\chi}}^4}{m_{\widetilde{t}}^5}\int^{(m_{\widetilde{t}}-m_q)^2}_{m_{\widetilde{\nu}}^2} d P_{\widetilde{\chi}}^2 \frac{ m_q m_{\widetilde{\chi}} \sqrt{\bigg(\frac{m_{\widetilde{t}}^2+P_{\widetilde{\chi}}^2-P_{\widetilde{\chi}}^2}{2m_{\widetilde{t}}}\bigg)^2-P_{\widetilde{\chi}}^2}(P_{\widetilde{\chi}}^2-m_{\widetilde{\nu}}^2 )^2}{P_{\widetilde{\chi}}^2(P_{\widetilde{\chi}}^2-m_{\widetilde{\chi}}^2)^2}~.~
\end{align}}In Eq.~(\ref{gammastop}), one can easily track the origin of each term, back to the chiral structure of the vertices. We can express the effective coupling constants in terms of the fundamental coupling constants and mixing angles. For $\widetilde{t}\rightarrow t\nu\widetilde{\nu}$, we have~\cite{Rosiek:1995kg}
{\small\begin{align}
g_L^q=&\frac{-2\sqrt{2}i e}{3c_W}N_{11}C_{\widetilde{t}R}+ i y_t C_{\widetilde{t}L} N_{14},\quad 
g_R^q=\frac{i e}{\sqrt{2}s_Wc_w} C_{\widetilde{t}L}\L\frac{1}{3}N_{11}s_W+N_{12}c_W\R+i y_t N_{14}C_{\widetilde{t}R},\cr
g_L^l=&\frac{i e}{\sqrt{2}s_Wc_w} C_{\widetilde{\nu}L}\L N_{11}s_W+N_{12}c_W\R+i y_{\nu} N_{14} C_{\widetilde{\nu}R},\quad
g_R^l=i y_{\nu} N_{14} C_{\widetilde{\nu}L}~,~
\end{align}}where $C_{\widetilde{t}L}$ and $C_{\widetilde{t}R}$ respectively denote the left-  and right-handed stop components, and $C_{\widetilde{\nu}L}$ and $C_{\widetilde{\nu}L}$ respectively denotes the left- and right-handed sneutrino components.
Each of them has two parts, the bino/wino and Higgsino contributions. For $\widetilde{t}\rightarrow bl\widetilde{\nu}$,
\begin{align}
g_L^q=&iy_b C_{\widetilde{t}L}U_{12},\quad
g_R^q=\frac{i e}{s_W} C_{\widetilde{t}L} V_{11}+iy_t C_{\widetilde{t}R} V_{12} ~,~\cr
g_L^l=&\frac{i e}{s_W}C_{\widetilde{\nu}L} V_{11} +iy_{\nu}C_{\widetilde{\nu}L} V_{12},\quad
g_R^l=i y_l C_{\widetilde{\nu}R} U_{12}~.~
\end{align}

The two-body decay processes of chargino/neutralino are shown in the middle of Fig.~\ref{procs}. The decay width is calculated to be
\begin{align}
\Gamma=\frac{(g_L^l)^2+(g_R^l)^2}{32 \pi}\frac{(m_{C/\chi}^2-m_{\widetilde{\nu}}^2)^2}{m_{C/\chi}^3}~,~
\end{align}
where $g_L^l$ and $g_R^l$ come from the vertex $g_L^l P_L+g_R^l P_R$.
For Higgsino (chargino) decay, $g_L^l$ and $g_R^l$ can be written as
\begin{align}
g_L^l=&i y_{\nu} C_{\widetilde{\nu}R}(i y_{l} C_{\widetilde{\nu}L}),\quad
g_R^l=i y_{\nu} C_{\widetilde{\nu}L}(i y_{\nu} C_{\widetilde{\nu}R})~.~
\end{align}

The three-body decay of chargino into neutralino and off-shell $W$ boson is shown in the right of Fig.~\ref{procs}. The decay width is given by
\begin{align}
\Gamma \times {\rm {Br}}(e\nu)=&\frac{1}{256\pi^3}\frac{m_{C}^5}{m_W^4}((g_L^W)^2g_2^2+(g_R^W)^2g_2^2)\widehat{f_{1}}
        +\frac{1}{128\pi^3}\frac{m_{C}^5}{m_W^4}(g_L^Wg_R^Wg_2^2)\widehat{f_{2}}~,~
\end{align}
where the integral functions 
{\small\begin{align}
\widehat{f_{1}}&=\frac{m_W^4}{m_{C}^8}\int^{(m_{C}-m_{\chi})^2}_0 d P_W^2\frac{\sqrt{k}}{(P_W^2-m_W^2)^2}\bigg(\frac{1}{2}(m_{C}^2-m_{\chi}^2)(m_{C}^2-m_{\chi}^2-P_W^2)-\frac{1}{6}k\bigg)~,~\cr
\widehat{f_{2}}&=\frac{m_W^5}{m_{C}^7}\int^{(m_{C}-m_{\chi})^2}_0 d P_W^2\frac{\sqrt{k}}{(P_W^2-m_W^2)^2} P_W^2~.~
\end{align}}In the Higgsino-limit, $k$ and $g_L^W, g_R^W, g_2$ are explicitly given by
\begin{align}
k&=m_{C}^4+m_{\chi}^4+P_W^4-2(m_{C}^2m_{\chi}^2+P_W^2m_{C}^2+P_W^2m_{\chi}^2)~,~\cr
g_L^W&=g_R^W=g_2=\frac{ie}{\sqrt{2}s_W}~.~
\end{align}

\section{Simulation and Validation}\label{vali}

In this appendix we give some details about the simulation for signatures used in the text. To ensure the validation of our simulation, we also presented the check for the usual SUSY searches in the relevant literatures.

Without loss (much) of generality, in the simulation only one specific chiral structure of the simplified model given in Eq.~(\ref{simp})  is considered. Concretely, we  work in the  Higgsino-limit and take $\tilde{t}_L \to t_R \tilde{H}^{0*}_u \to t_R (\nu_L \tilde{\nu}_R)$ for top1/3,  and $\tilde{t}_R \to b_L \tilde{H}^{+*}_{u} \to b_L (e_L \tilde{\nu}_R)$ for top2/4. The UFO  model files of the corresponding simplified model is generated by SARA4.0~\cite{Staub:2013tta}. Practically, we employ decays at the matrix element level so as to gain full helicity information in the final states, and find that decays in Pythia, which disregards helicity, leads to almost the same result. This justifies our simplifying treatment at the beginning.



Comments are in orders. Firstly, we do not discriminate electron and muon, because the difference between their detector efficiency is within the uncertainty in the simulation. Thus, as long as the flavour of the final state lepton is not concerned, which is the case in the current experimental analysis, our simulation is applicable to both flavour of the sneutrino LSP. Secondly, we emphasized that that the default PGS does not identify the sneutrino LSP as missing energy, and consequently it might be used to reconstruct jet. We have adapted the PGS for all flavors of sneutrinos being recognized as missing energy.


\begin{table}[htb]
\begin{center}
\begin{tabular}{|c|c|c|c|c|} \hline
\multicolumn{5}{|c|}{Low $\Delta$M} \\ \hline
Signal Region & $E^{\text{miss}}_T>100$ GeV & $E^{\text{miss}}_T>150$ GeV & $E^{\text{miss}}_T>200$ GeV & $E^{\text{miss}}_T>250$ GeV \\ \hline
Background & 1662 $\pm$ 203 & 537 $\pm$ 75 & 180 $\pm$ 28 & 66 $\pm$ 13 \\ \hline
Data & 1624 & 487 & 151 & 52 \\ \hline
$N_{up}$ & 473 & 189 & 59 & 29 \\ \hline \hline
\multicolumn{5}{|c|}{High $\Delta$M} \\ \hline
Signal Region & $E^{\text{miss}}_T>100$ GeV & $E^{\text{miss}}_T>150$ GeV & $E^{\text{miss}}_T>200$ GeV & $E^{\text{miss}}_T>250$ GeV \\ \hline
Background & 79 $\pm$ 12 & 38 $\pm$ 7 & 19 $\pm$ 5 & 9.9 $\pm$ 2.7 \\ \hline
Data & 90 & 39 & 18 & 5 \\ \hline
$N_{up}$ & 46 & 22 & 15 & 7.1 \\ \hline
\end{tabular}
\end{center}
\caption{CMS-PAS-SUS-13-011.}
\label{c011up}
\end{table}

\begin{figure}[htb]
\begin{center}
\includegraphics[width=0.45\columnwidth]{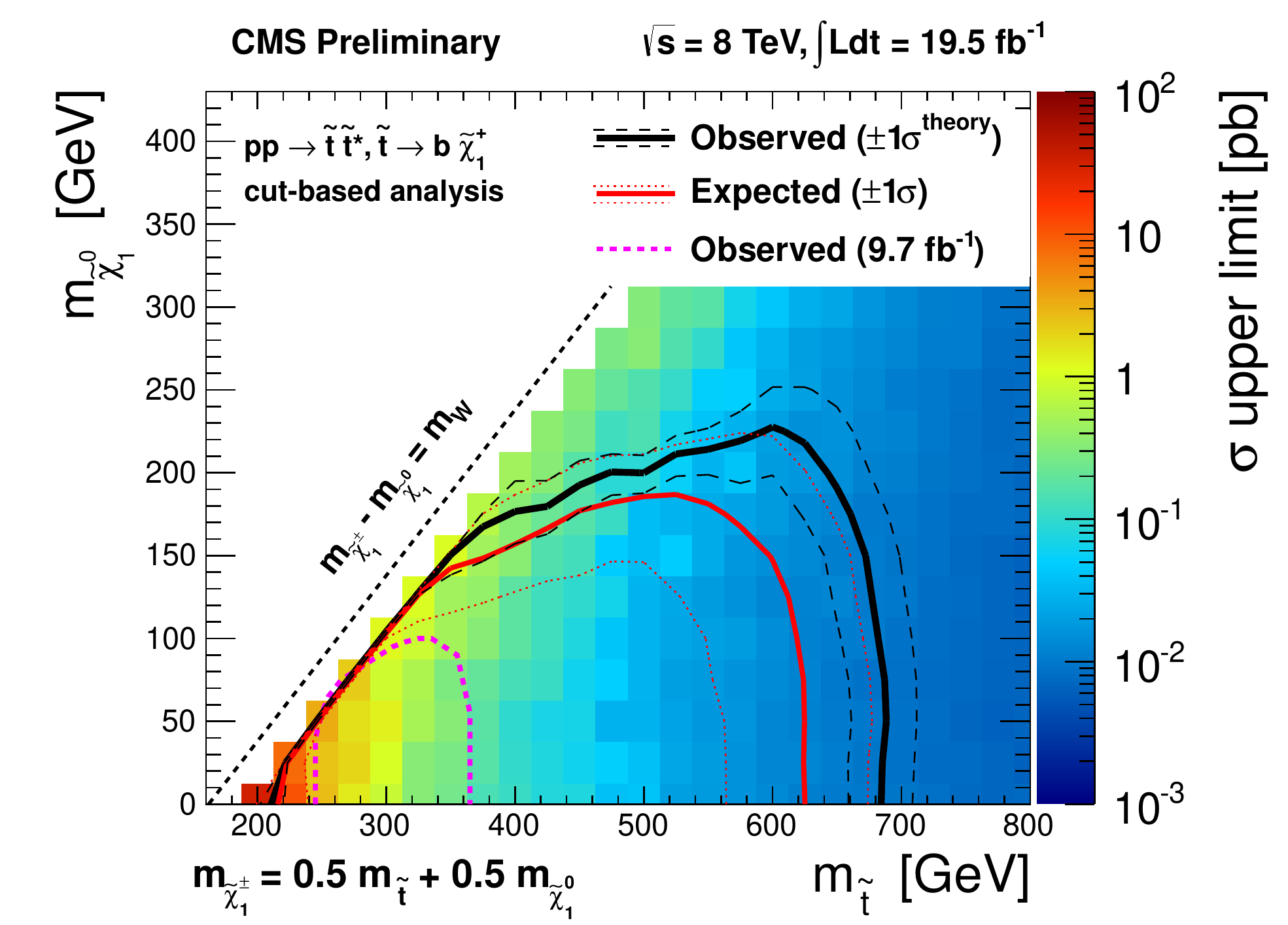}
\includegraphics[width=0.5\columnwidth]{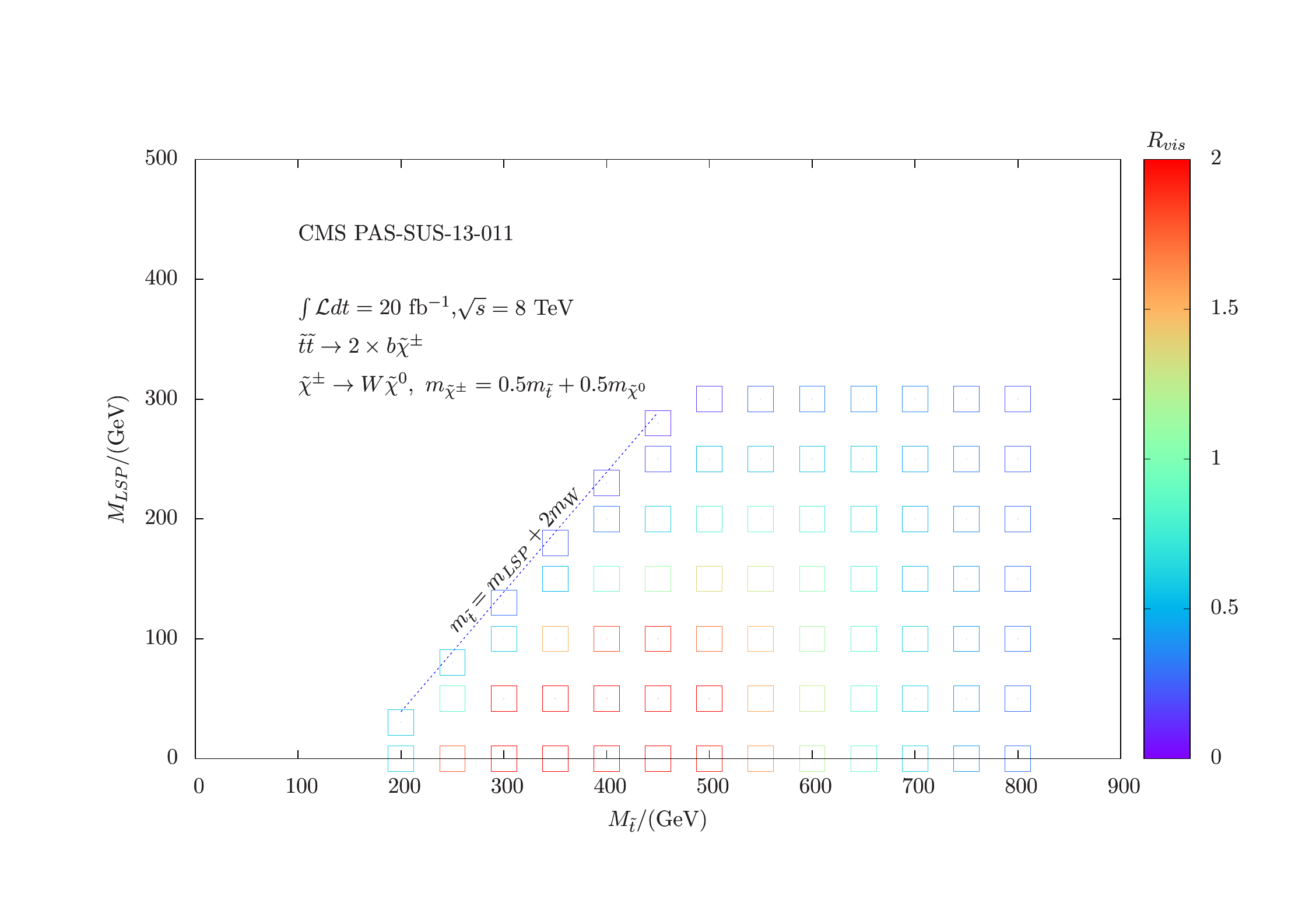}
\caption{Left: Experimental bounds on masses of stop and neutralino LSP given in \cite{CMS-PAS-SUS-13-011}.  Right: Our check. Color scheme is the same as Fig.~\ref{emusv}.\label{c011ck}}
\end{center}
\end{figure}

We now proceed to discuss the validation of our Monte Carlo simulation. We examined all the channels which give the strongest bounds in the text and found that the results are in accord with the LHC results, with the maximal deviation less than 50 GeV. For illustration, we show the check on Ref.~\cite{CMS-PAS-SUS-13-011}. It searchs the final states with one isolated lepton, $b$-jets and missing energy. This search is sensitive to both $\tilde{t} \to t \nu \tilde{\nu}$ channel and $\tilde{t} \to b \tau \nu_{\tau}$ channel. 
In order to check both the Bayesian procedure used to estimate the upper limit and the cuts implemented, we generate the same process as that of Ref.~\cite{CMS-PAS-SUS-13-011}. We scan the parameters $m_{\tilde{t}}$, $m_{\tilde{\chi}^{\pm}}$ and $m_{\rm{LSP}}$ with $m_{\tilde{\chi}^{\pm}_1} = 0.5 m_{\tilde{t}} + 0.5 m_{\rm{LSP}}$ and then compare our bounds with theirs. The results are shown in Fig.~\ref{c011ck}. In Table~\ref{c011up}, the upper limit of number of events for new physics contribution in each signal regions is given.

\bibliography{sneutrino}
\bibliographystyle{utphys}

\end{document}